\documentclass[journal]{IEEEtran}
\usepackage{bbm}
\usepackage{cite}
\usepackage{amsfonts}
\usepackage{mathrsfs}
\usepackage{graphicx}
\usepackage{amssymb}
\usepackage{latexsym}
\usepackage{amsmath}
\usepackage{stfloats}
\usepackage{cases}
\usepackage{setspace}
\usepackage{bigstrut}
\usepackage{bm}
\usepackage{url}
\usepackage{color}
\usepackage{algorithm}
\usepackage{algorithmic}

\newtheorem{property}{Property}

\begin{document}

\title{Deep Learning Based FDD Non-Stationary Massive MIMO Downlink Channel Reconstruction}

\author{Yu~Han, Mengyuan~Li,~\IEEEmembership{Student Member,~IEEE,}
Shi~Jin,~\IEEEmembership{Senior Member,~IEEE,}\\
Chao-Kai~Wen,~\IEEEmembership{Member,~IEEE,}
and~Xiaoli~Ma,~\IEEEmembership{Fellow,~IEEE}
\thanks{Y. Han, M. Li and S. Jin are with the National Mobile Communications Research Laboratory, Southeast University, Nanjing, 210096, P. R. China (email: hanyu@seu.edu.cn, mengyuanli@seu.edu.cn, and jinshi@seu.edu.cn).}
\thanks{C.-K. Wen is with the Institute of Communications Engineering, National Sun Yat-sen University, Kaohsiung 804, Taiwan (e-mail: chaokai.wen@mail.nsysu.edu.tw).}
\thanks{X. Ma is with the School of Electrical and Computer Engineering, Georgia Institute of Technology, Atlanta, GA 30332, USA (email: xiaoli@gatech.edu).}}

\maketitle

\begin{abstract}
This paper proposes a model-driven deep learning-based downlink channel reconstruction scheme for frequency division duplexing (FDD) massive multi-input multi-output (MIMO) systems. The spatial non-stationarity, which is the key feature of the future extremely large aperture massive MIMO system, is considered. Instead of the channel matrix, the channel model parameters are learned by neural networks to save the overhead and improve the accuracy of channel reconstruction. By viewing the channel as an image, we introduce You Only Look Once (YOLO), a powerful neural network for object detection, to enable a rapid estimation process of the model parameters, including the detection of angles and delays of the paths and the identification of visibility regions of the scatterers. The deep learning-based scheme avoids the complicated iterative process introduced by the algorithm-based parameter extraction methods. A low-complexity algorithm-based refiner further refines the YOLO estimates toward high accuracy. Given the efficiency of model-driven deep learning and the combination of neural network and algorithm, the proposed scheme can rapidly and accurately reconstruct the non-stationary downlink channel. Moreover, the proposed scheme is also applicable to widely concerned stationary systems and achieves comparable reconstruction accuracy as an algorithm-based method with greatly reduced time consumption.
\end{abstract}


\section{Introduction}\label{Sec:Introduction}

The acquisition of downlink channel state information (CSI) in frequency division duplex (FDD) massive multi-input multi-output (MIMO) systems has been a long-term problem that obsesses the mobile communication industry \cite{Araujo2016,Elijah2016,Fan2017}. Without the reciprocity between uplink and downlink, the downlink CSI has to be obtained through downlink training and feedback, causing a large amount of overhead. Recently, studies have suggested to utilize the spatial reciprocity \cite{Hugl2002} to reduce the cost of downlink CSI acquisition. Uplink and downlink channels has a similar spatial domain given that they share the same space and scatterers. Thereafter, part of downlink CSI can be derived from the uplink CSI.

\subsection{Related work}

Many related works have been developed to estimate or reconstruct the FDD massive MIMO downlink channels under different channel models. For clustering channels, where a continuous spatial region has distinct power, the correlation matrix is introduced to describe the power distribution of the channel in the spatial domain \cite{Xie2018,Haghighatshoar2018,Khalilsarai2018}. The downlink correlation matrix can be derived from the uplink, and only the downlink instantaneous CSI should be estimated in the downlink. For limited scattering channels, the angle and delay of each propagation path are common in uplink and downlink, and only the downlink gains should be estimated in the downlink \cite{Zhang2018,Han2019TWC,Han2019TCOM,Han2019JSTSP}. These spatial reciprocity-based methods effectively ease the burden of downlink training and feedback; they have great potential in the future use. The spatial reciprocity does not indicate that the downlink CSI can be completely derived from the uplink. The overhead of downlink training and feedback is still required.

In recent years, the rapid development of deep learning techniques stimulates their wide applications to various areas, including localization \cite{Ihsan2018} and FDD downlink channel estimation or prediction \cite{Alrabeiah2019,Arnold2019,Safari2018,Wang2019UL,Yang2019,Liu2019,Dong2018}. Most of these methods are based on an assumption that a mapping function exists between the uplink and the downlink channels, which can be conveniently learned by deep networks instead of traditional algorithms. With the mapping function, the downlink channel matrix can be directly predicted from the uplink channel matrix \cite{Alrabeiah2019,Arnold2019,Safari2018,Wang2019UL,Yang2019}.
The channel matrix of the massive MIMO multicarrier system also can be illustrated by an image. The channel as an image is an interesting strategy \cite{Wen2018,Wang2019Deep}, which enables the application of advanced deep learning-based image processing methods. For the downlink channel prediction problem, the uplink and downlink channel images are stacked together into a large image \cite{Safari2018}. With the uplink channel, the base station (BS) draws one half of the image. The other half of the image, which is currently white, is the downlink channel to be predicted. The downlink channel prediction method works as a painter to complete the other half of the image through image processing methods, such as generative adversarial networks. The methods in \cite{Alrabeiah2019,Arnold2019,Safari2018,Wang2019UL,Yang2019} do not require feedback, thereby raising the interests of the industry. However, the assumption of the mapping function between the uplink and the downlink channels is invalid in complicated multipath propagation scenarios \cite{Han2019TWC}.

To address this problem, the downlink channel is estimated with minor feedback in \cite{Liu2019}. The downlink channel matrix obtained at the user side is initially encoded, sent to the BS, and then decoded with the aid of uplink channel matrix. Besides, the downlink subchannel on subarray $\mathcal{A}$ (denoted by ${\bf H}_\mathcal{A}$) is correlated with that on subarray $\mathcal{B}$ (denoted by ${\bf H}_\mathcal{B}$) when spatial stationarity exists. If ${\bf H}_\mathcal{B}$ is obtained, then ${\bf H}_\mathcal{A}$ can be learned from ${\bf H}_\mathcal{B}$, with the cost of downlink training and feedback overhead to acquire ${\bf H}_\mathcal{B}$ \cite{Dong2018}. These methods are applicable in practice. However, they ignore the channel model and directly predicts the channel matrix. Under multipath propagation conditions, the accuracy of estimation is affected if the compression rate is low or the scale of subarray $\mathcal{B}$ is much smaller than that of subarray $\mathcal{A}$. On the contrary, scaling up the compression rate or the size of subarray $\mathcal{B}$ further increases the overhead amount. This contradiction limits the performance of these data-driven methods. Therefore, referring to the channel model is necessary to increase the efficiency of deep learning-based channel estimation.

In massive MIMO systems, where the scale of antenna array is extremely large, the signal reflected by a scatter does not arrive at the entire array, and the channel begins to show spatial non-stationarity \cite{Carvalho2019,Ali2019,Amiri2018}. Non-stationarity is a distinct feature in future massive MIMO systems, where the array may be widely spread on the wall of a building. The non-stationary channel is more complicated than the stationary ones because the visibility region of each scatterer, that is, the part of the array that can receive signals from the scatterer, should also be considered in the channel model. Thus, estimating the downlink channel of an FDD non-stationary massive MIMO system is challenging. The methods in \cite{Xie2018,Haghighatshoar2018,Khalilsarai2018,Zhang2018,Han2019TWC,Han2019TCOM,Alrabeiah2019,Arnold2019,Safari2018,Wang2019UL,Yang2019,Liu2019,Dong2018} are designed for FDD stationary massive MIMO systems. The study on the estimation of FDD non-stationary massive MIMO downlink channels is limited.

\subsection{Contribution of this paper}

We focus on the downlink channel reconstruction of FDD non-stationary massive MIMO systems. In accordance with the multipath channel model, the downlink channel can be reconstructed by the downlink gains, angles, delays, and visibility regions of the propagation paths. We acquire the frequency-independent parameters, including the angles, delays, and visibility regions, from the uplink and then estimate the downlink gains from the downlink given the spatial reciprocity.
If we apply the iteration-based algorithms to estimate the frequency-independent parameters, then the complexity of algorithm explosively increases. To tackle this problem, we propose a model-driven deep learning-based downlink channel reconstruction scheme, which has the following advantages.

\subsubsection{Power of using You Only Look Once (YOLO)}
YOLO, a fast object detection neural network that detects all the objects by looking at the image for only once, can effectively tackle the problem of explosive complexity. We introduce YOLO to detect each path with much reduced processing time compared with using iteration-based algorithms \cite{Han2019TWC}. With the bounding boxes designed in this study, the frequency-independent parameters of each path can be conveniently obtained.

\subsubsection{Efficiency of model-driven deep learning}
We do not follow the data-driven methods to learn the downlink channel matrix, but we learn the parameters of the paths in the channel. Driven by the channel model, the number of coefficients to be estimated is much smaller than the data-driven methods. Accordingly, the downlink training and feedback overhead is greatly reduced, and the accuracy of the reconstruction is guaranteed.

\subsubsection{Ability to identify visibility regions}
The visibility region of each scatterer consists of one or several subarrays that receive the signal reflected by the scatterer. Two algorithms are proposed to identify the visibility regions in different approaches. Either approach achieves a successful ratio of more than 98\%.

\subsubsection{Refinement of estimates}
A low-complexity refinement module is introduced to refine the estimates of angles and delays. After refinement, the normalized mean square error (NMSE) of uplink channel reconstruction is reduced from $-8$ dB to ${-28}$ dB at signal-to-noise ratio (SNR) $=0$ dB.

\subsubsection{Applicability to stationary cases}
The proposed scheme also works in the reduced FDD stationary massive MIMO systems, where most existing works focus on, and the NMSE performance is close to that of the algorithm-based reconstruction. On the basis of the stationarity of each subarray, an alternative scheme for non-stationary systems is further formulated and evaluated. Nevertheless, the proposed scheme is proven to be more efficient than the alternative one.

In the following section, we initially introduce the system model. The rationale of deep learning-based parameter estimation and the working steps of the proposed scheme are provided in Sections \ref{Sec:YOLO} and \ref{Sec:scheme}, respectively. Section \ref{Sec:simulations} evaluates the scheme and Section \ref{Sec:conclusion} concludes the paper.

\emph{Notations}---We denote scalars by letters in normal fonts, and use uppercase and lowercase boldface letters to represent matrices and vectors, respectively. The superscripts $(\cdot)^*$, $(\cdot)^{T}$, and $(\cdot)^{H}$ indicate conjugate, transpose, and conjugate transpose, respectively. $\mathbb{E}\{\cdot\}$ means considering the expectation with respect to the random variables inside the brackets. $\odot$ and $\otimes$ denote taking the Hadamard and Kronecker products, respectively. We also denote the absolute value and modulus operations by $\left| \cdot \right|$ and $\left\| \cdot \right\|$ and use $\left\lfloor \cdot \right\rfloor$ and $\left\lceil \cdot \right\rceil$ to round a decimal number to its nearest lower and higher integers, respectively. $[{\bf A}]_{i,:}$, $[{\bf A}]_{:,j}$, and $[{\bf A}]_{i,j}$ represent the $i$th row, the $j$th column, and the $(i,j)$th entry of matrix $\bf A$.

\section{System Model}\label{Sec:SystemModel}

In a cell of the FDD massive MIMO system, the BS is located at the cell center and equipped with an $M$-element uniform linear array (ULA), where $M$ is large. The distance between two adjacent ULA elements is $d$. Single-antenna users are randomly distributed in the cell. The reconstruction of each user channel is conducted independently; therefore, we focus on a single user.

The system works in the FDD duplexing mode. The uplink and downlink carrier frequencies are $f^{\rm ul}$ and $f^{\rm dl}$, respectively. The uplink and downlink carrier wavelengths are approximately equal and unified as $\lambda$ given $|f^{\rm ul}-f^{\rm dl}|\ll f^{\rm ul}$, $f^{\rm dl}$. Orthogonal frequency division multiplexing (OFDM) is applied. The area band has $N$ subcarriers with spacing $\Delta f$ between two adjacent subcarriers.

\subsection{Non-stationarity}

\begin{figure}
  \centering
  \includegraphics[scale=0.55]{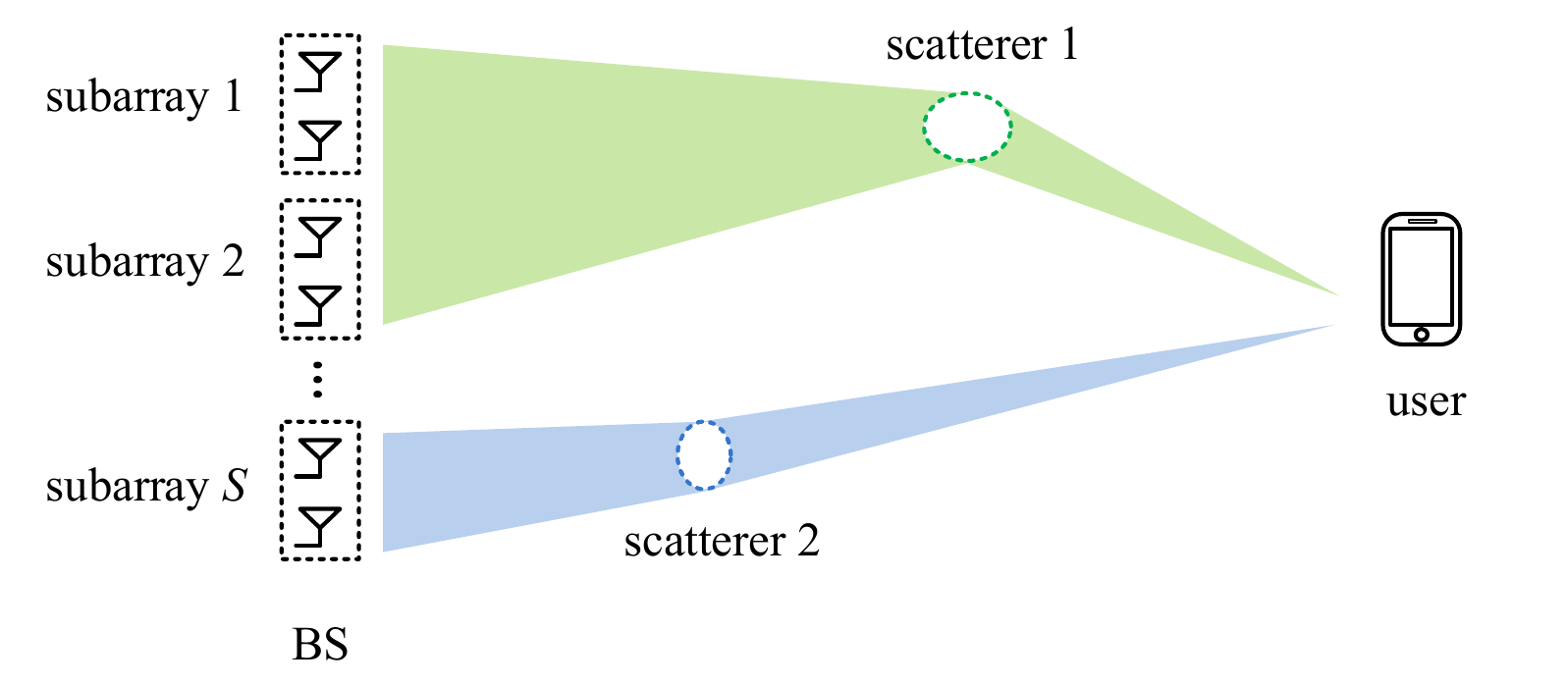}
  \caption{Spatial non-stationarity. Path 1 arrives at subarrays 1 and 2, whereas path 2 arrives at subarray $S$.} \label{Fig:NonStationary}
\end{figure}

The channel between the BS and the user comprises $L$ paths, corresponding to $L$ scatterers. For the line-of-sight path, the scatterer is the user antenna itself. The ULA at BS experiences spatial non-stationarity due to the large aperture of the array. Signals reflected by a scatterer may arrive at the entire ULA or a part of the ULA, as shown in Fig.~\ref{Fig:NonStationary}.

The ULA is uniformly segmented into $S$ subarrays, each with $M/S$ elements.
The set of adjacent subarrays that can see scatterer $l$ is defined as the visibility region of scatterer $l$, denoted as follows:
\begin{equation}\label{Eq:VRscatterer}
\Phi_l = \{ s_{l,{\rm start}},s_{l,{\rm start}}+1,\ldots,s_{l,{\rm end}}-1,s_{l,{\rm end}}\},
\end{equation}
where $s_{l,{\rm start}}$ and $s_{l,{\rm end}}$ are the first and last subarrays that can receive signals reflected from scatterer $l$, respectively, satisfying $1\le s_{l,{\rm start}}\le s_{l,{\rm end}} \le S$.

Similarly, the visibility region of subarray $s$ includes the scatterers that can see subarray $s$, denoted as follows:
\begin{equation}\label{Eq:VRsubarray}
\Psi_s = \{ l_{s,1},l_{s,2},\ldots,l_{s,L_s} \},
\end{equation}
where $1\le l_{s,i} \le L$ holds for $i=1,\ldots,L_s$ and $L_s$ is the number of scatterers that can reflect signals to subarray $s$, satisfying $0\le L_s\le L$.

The example of Fig.~\ref{Fig:NonStationary} is considered to illustrate the visibility regions. For the scatterers, $\Phi_1=\{1,2\}$ and $\Phi_2=\{S\}$, and for the subarrays, $\Psi_1=\Psi_2=\{1\}$ and $\Psi_S=\{2\}$.

\subsection{Spatial reciprocity}

Although the uplink and downlink channels are in different frequency bands, they share the space and the scatterers. On the basis of this spatial reciprocity, the delay and angle of the $l$ path, as well as the visibility region of the $l$th scatterer, are frequency-independent and identical in uplink and downlink. We denote $\tau_l$ and $\theta_l$ as the delay and angle of the $l$ path, respectively, satisfying\footnote{In practical systems, $\tau_l$ should be not greater than the cyclic-prefix length. Here, we relax this restriction by assuming that the cyclic-prefix length is equal to the symbol length.} $0\le\tau_l\le {1}/{\Delta f}$ and $0\le\theta_l\le 2\pi$. The frequency-independent parameters are $\tau_l$, $\theta_l$, and $\Phi_l$, where $l=1,\ldots,L$.

When reflection or scattering occurs, the phase shift amount differs in the uplink and downlink due to different carrier frequencies. Consequently, the complex gains are frequency-dependent and different in uplink and downlink. We denote $g^{\rm ul}_l$ and $g^{\rm dl}_l$ as the uplink and downlink complex gains of the $l$th path, respectively, which are different from each other.

\subsection{Channel model}

In the baseband, the frequency of the first subcarrier of the downlink OFDM module is regarded as 0 Hz, and that of the uplink OFDM module is $f^{\rm ul}-f^{\rm dl}$. The non-stationary downlink channel between the BS and the user $k$ across all antennas and subcarriers is modeled as
\begin{equation}\label{Eq:DLchannel}
{\bf H}^{\rm dl} = \sum_{l=1}^{L} g^{\rm dl}_l \left({\bf a}(\Theta_l) \odot {\bf p}(\Phi_l)\right) {\bf q}^T(\Gamma_l),
\end{equation}
where ${\bf H}^{\rm dl}\in\mathbb{C}^{N \times M}$ is in the antenna subcarrier domain,
\begin{equation}\label{Eq:GammaTheta}
\Theta_l=\frac{d}{\lambda}\sin\theta_l, \quad \Gamma_l=\Delta f \tau_l,
\end{equation}
simplify the expressions and have frequency-independency, satisfying $0\le\Theta_l\le 1$ and $0\le\Gamma_l\le 1$,
\begin{equation}\label{Eq:avec}
{\bf a}(\Theta)=\left[1, e^{j2\pi\Theta}, \ldots, e^{j2\pi(M-1)\Theta}\right]^T
\end{equation}
is the steering vector of the ULA, ${\bf p}(\Phi) \in \mathbb{Z}^{M\times 1}$ selects the ULA elements that are in the subarrays in $\Phi$, the $m$th entry is
\begin{equation}\label{Eq:pvec}
\left[{\bf p}(\Phi)\right]_m =
\begin{cases}
1, & \text{if $\lceil \frac{mS}{M}\rceil \in \Phi$,} \\
0, & \text{else,}
\end{cases}
\end{equation}
and
\begin{equation}\label{Eq:qvec}
{\bf q}(\Gamma)=\left[1, e^{j2\pi\Gamma}, \ldots, e^{j2\pi(N-1)\Gamma}\right]^T
\end{equation}
is the phase shift vector across the OFDM subcarriers.

Given the spatial reciprocity, the uplink baseband channel is expressed as
\begin{equation}\label{Eq:ULchanneltmp}
{\bf H}^{\rm ul} = \sum_{l=1}^{L} g^{\rm ul}_l e^{j2\pi\left(f^{\rm ul}-f^{\rm dl}\right)\tau_l} \left({\bf a}(\Theta_l) \odot {\bf p}(\Phi_l)\right) {\bf q}^T(\Gamma_l),
\end{equation}
where ${\bf H}^{\rm ul}\in\mathbb{C}^{N \times M}$ is in the antenna subcarrier domain. We further define
\begin{equation}\label{Eq:ULegain}
\alpha_l = g^{\rm ul}_l e^{j2\pi\left(f^{\rm ul}-f^{\rm dl}\right)\tau_l}
\end{equation}
as the effective uplink gain of the $l$th path and simplify the uplink channel model \eqref{Eq:ULchanneltmp} as
\begin{equation}\label{Eq:ULchannel}
{\bf H}^{\rm ul} = \sum_{l=1}^{L} \alpha_l \left({\bf a}(\Theta_l) \odot {\bf p}(\Phi_l)\right) {\bf q}^T(\Gamma_l).
\end{equation}

\section{Acquire model parameters through learning}\label{Sec:YOLO}

We focus on the reconstruction of the downlink channel ${\bf H}^{\rm dl}$, which is a fundamental requirement to harvest the spatial multiplexing gain of FDD massive MIMO downlink. Given the channel model \eqref{Eq:DLchannel}, we can reconstruct the downlink channel with the model parameters, i.e., $\Theta_l$, $\Gamma_l$, $\Phi_l$, and $g^{\rm dl}_l$ of each path. Thereafter, the acquisition of these model parameters becomes the primary task of downlink channel reconstruction. Notably, the number of paths (i.e., $L$) is also unknown.

\begin{figure*}
  \centering
  \includegraphics[scale=1]{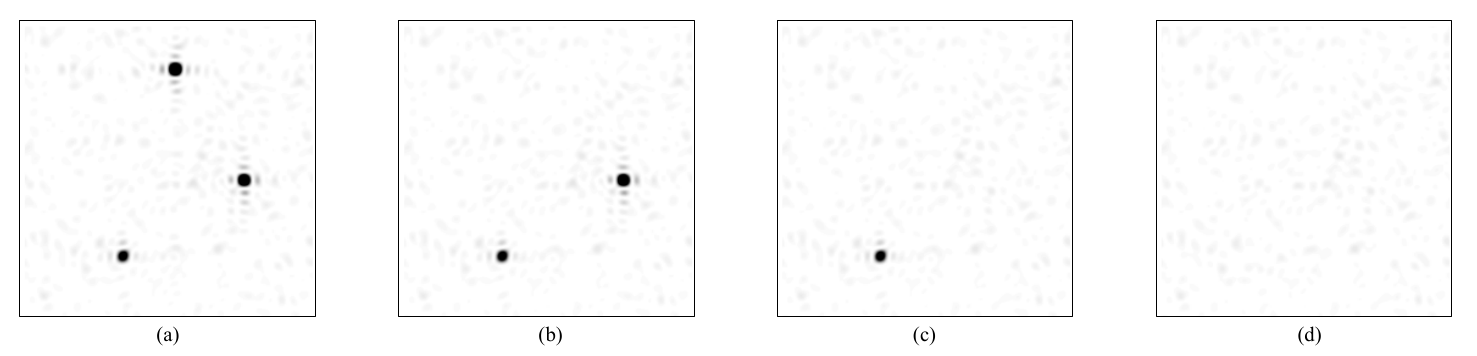}
  \caption{Residues of pilots after each iteration of the NOMP algorithm, where the horizontal and vertical axes represent delay and angle, respectively. Only one path is detected at each iteration. If three paths exist, then the NOMP algorithm requires three iterations to find all the paths.}\label{Fig:NOMPprocess}
\end{figure*}

On the basis of the spatial reciprocity, the model parameters are divided into two categories, that is, the frequency-independent parameters (i.e., $\Theta_l$, $\Gamma_l$, and $\Phi_l$) and the frequency-dependent parameters (i.e., $g^{\rm dl}_l$). We estimate the frequency-independent parameters in the uplink and acquire the frequency-dependent parameters through downlink training and feedback \cite{Han2019TWC}. This method greatly relaxes the overhead requirement on the downlink training and reduces the feedback amount from $MN$ to $L$ complex numbers compared with traditional linear channel estimation methods, such as least squares (LS) and linear minimum mean square error (LMMSE) estimators, which can also be regarded as data-driven methods.

The frequency-independent parameters are estimated during the uplink sounding phase.
The uplink all-one pilots received by the BS across all antennas and subcarriers are expressed as
\begin{equation}\label{Eq:ULpilots}
{\bf Y}^{\rm ul} = \sqrt{P^{\rm ul}} \sum_{l=1}^{L} \alpha_l \left({\bf a}(\Theta_l) \odot {\bf p}(\Phi_l)\right) {\bf q}^T(\Gamma_l) + {\bf Z}^{\rm ul},
\end{equation}
where ${\bf Y}^{\rm ul}\in\mathbb{C}^{M\times N}$ is in the antenna subcarrier domain, $P^{\rm ul}$ is the transmitted power of user, and ${\bf Z}^{\rm ul}\in\mathbb{C}^{M\times N}$ is the uplink complex Gaussian noise whose elements are independent and identically distributed (i.i.d.) with zero mean and unit variance. ${\bf Y}^{\rm ul}$ is a noisy mixture composed of the pilot components that travel along the $L$ paths and the additive Gaussian noise. We aim to extract $\Theta_l$, $\Gamma_l$, and $\Phi_l$ from the noisy mixture ${\bf Y}^{\rm ul}$.

This section formulates two key problems that lie in the extraction of these frequency-independent parameters through reviewing the authors' previous work \cite{Han2019TWC} for FDD stationary massive MIMO systems, and then introduces deep learning to tackle these problems.

\subsection{Problem formulation}

Newton orthogonal matching pursuit (NOMP) algorithm \cite{Mamandipoor2016} is adopted in \cite{Han2019TWC} to estimate $\Theta_l$, $\Gamma_l$, and $\alpha_l$ from ${\bf Y}^{\rm ul}$ successively. In the $l$th iteration, NOMP estimates $\Theta_l$ and $\Gamma_l$ of the $l$th path and removes the pilot component along this path from ${\bf Y}^{\rm ul}$. The residues of ${\bf Y}^{\rm ul}$ after each iteration are illustrated in Fig.~\ref{Fig:NOMPprocess}, where $M=N=64$ and $L=3$. Figs.~\ref{Fig:NOMPprocess}(a) and (b) show that the NOMP algorithm can recognize only the pilot component with the largest power. Subsequently, this strongest component is removed and the updated residue contains two components. After three iterations, all the components are removed, that is, all the paths are detected, thereby leaving only the noise in the residue, as illustrated in Fig.~\ref{Fig:NOMPprocess}(d). The iteration-based algorithm requires $L$ rounds of detection to recognize all the paths. The complexity of the NOMP algorithm is $\mathcal{O}(LMN\log(MN))$. If $M$, $N$, and $L$ grow large, then the processing time is considerably long. To avoid the latency caused by using high-complexity algorithms, we raise the first question as follows:
\begin{itemize}
  \item {\bf Q1}: Can we rapidly recognize the angles and delays of all the paths?
\end{itemize}

The scheme proposed in \cite{Han2019TWC} was designed for spatially stationary systems and cannot identify $\Phi_l$, which is also frequency-independent in non-stationary systems. One solution is to estimate $\Phi_l$ together with $\Theta_l$ and $\Gamma_l$ at the $l$th iteration of the NOMP algorithm. With increasing parameters to be estimated, the computation complexity of the updated algorithm is further increased. All possible solutions of $\Phi_l$, which is further transformed to the search of $s_{l,{\rm start}}$ and $s_{l,{\rm end}}$, are exhaustively tested. The complexity of the updated algorithm is $1+2+\cdots+S=(S^2+S)/2$ times that of the NOMP algorithm, thereby resulting in incredibly long processing time, which is unacceptable in practice. Thus, we raise the second question as follows:
\begin{itemize}
  \item {\bf Q2}: How to efficiently identify the visibility regions?
\end{itemize}

\subsection{Sparse image of uplink pilots}

\begin{figure*}
  \centering
  \includegraphics[scale=0.6]{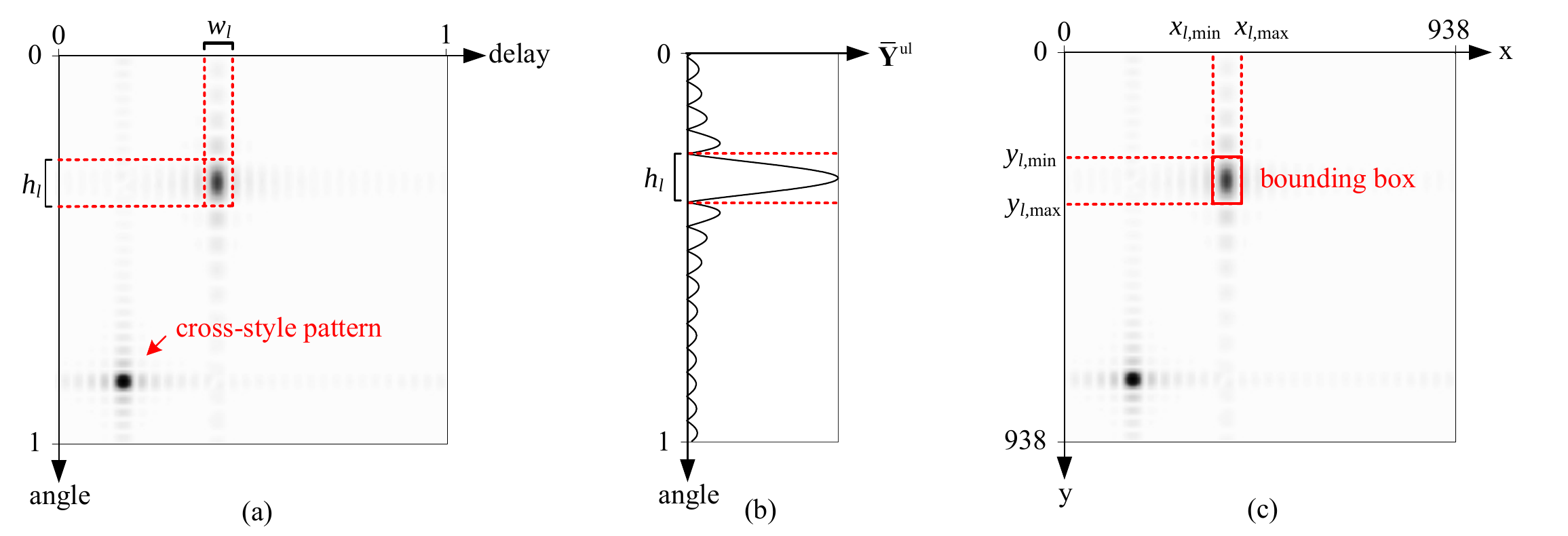}
  \caption{(a) Image of uplink pilots and the coordinate system of image. (b) Sinc-function pattern of a column of ${\bar{\bf Y}}^{\rm ul}$ and the width of a dark spot (suppose only one path exists). (c) Coordinate system of network and the bounding box. } \label{Fig:CoordinateSystems}
\end{figure*}

When observing Fig.~\ref{Fig:NOMPprocess}(a) which shows significant sparsity, we can rapidly determine the three paths in the channel. This process is fast without adopting iterations or generating figures of residues, which can imitated by artificial intelligence. Therefore, prior to answering the two questions, we initially investigate the sparse image of uplink pilots.

In massive MIMO OFDM systems, $L\ll MN$ typically holds. After transforming ${\bf Y}^{\rm ul}$ from the antenna subcarrier domain to the angular temporal domain, the pilots show sparsity. The angular and temporal transformation matrices are defined as
\begin{equation}\label{Eq:Uatrans}
{\bf U}_{\rm a}=\left[{\bf a}(0), {\bf a}\left(-\frac{1}{\gamma_{\rm a} M}\right), \ldots, {\bf a}\left(-\frac{\gamma_{\rm a}M-1}{\gamma_{\rm a}M}\right)\right]
\end{equation}
and
\begin{equation}\label{Eq:Uttrans}
{\bf U}_{\rm t}=\left[{\bf q}(0), {\bf q}\left(-\frac{1}{\gamma_{\rm t} N}\right), \ldots, {\bf q}\left(-\frac{\gamma_{\rm t}N-1}{\gamma_{\rm t}N}\right)\right]
\end{equation}
respectively, where $\gamma_{\rm a}$ and $\gamma_{\rm t}$ are oversampling rates.
Thereafter, the uplink received pilots in temporal angular domain are calculated as
\begin{equation}\label{Eq:ULtildePilot}
{\bar{\bf Y}}^{\rm ul} = {\bf U}^H_{\rm a}{\bf Y}^{\rm ul}{\bf U}_{\rm t},
\end{equation}
where ${\bar{\bf Y}}^{\rm ul}\in\mathbb{C}^{\gamma_{\rm a}M\times\gamma_{\rm t}N}$ is a sparse matrix.
We normalize the module of each entry of ${\bar{\bf Y}}^{\rm ul}$ and obtain a new real-valued matrix ${\tilde{\bf Y}}^{\rm ul}$, whose $(m,n)$th entry is
\begin{equation}\label{Eq:ULbarPilot}
[{\tilde{\bf Y}}^{\rm ul}]_{m,n} = \frac{\eta|[{\bar{\bf Y}}^{\rm ul}]_{m,n}|}{\max_{i=1,\ldots,\gamma_{\rm t}N,k=1,\ldots,\gamma_{\rm a}M}|[{\bar{\bf Y}}^{\rm ul}]_{i,k}|}.
\end{equation}
The maximal entry of ${\tilde{\bf Y}}^{\rm ul}$ is normalized by $\eta$. This normalization can avoid the wide color range of the images in an extremely high SNR regime.

The image of ${\tilde{\bf Y}}^{\rm ul}$ is drawn by MATLAB, as an example shown Fig.~\ref{Fig:CoordinateSystems}(a), where $M=64$, $N=32$, $S=4$, and $L=2$. In the image, the horizontal axis represents delay (i.e., $\Gamma$) ranging from 0 to 1, and the vertical axis represents angle (i.e., $\Theta$) ranging from 0 to 1. In the coordinate system of the image, the upper left vertex is the origin (0,0).

The image has $L$ cross-style patterns, each corresponding to a path. The darkness of the cross-style pattern is determined by the gain of the path. Each cross-style pattern is composed of a strong dark spot at the center and four dotted tails that stretch upwards, downwards, leftward, and rightwards. Each dark spot has a semi-square or semi-rectangular shape and holds the following two properties.

\begin{property}\label{Theo:lightSpot1}
In the coordinate system of the image, the coordinates of the center of the dark spot are exactly the delay and angle of the $l$th path, i.e., $(\Gamma_l,\Theta_l)$.
\end{property}

\begin{proof}
Refer to Appendix A.
\end{proof}

\begin{property}\label{Theo:lightSpot2}
The width $w_l$ and height $h_l$ of the $l$th dark spot in Fig.~\ref{Fig:CoordinateSystems}(a) are given as follows:
\begin{equation}\label{Eq:WidthHeight}
w_l = \frac{2}{N}, \quad h_l = \frac{2S}{\left( s_{l,{\rm end}}-s_{l,{\rm start}}+1 \right)M}.
\end{equation}
\end{property}

\begin{proof}
Refer to Appendix B.
\end{proof}

The proofs show that the cross-style pattern is resulted from the sinc-function pattern of $\bar{\bf Y}^{\rm ul}$. We extract one column of $\bar{\bf Y}^{\rm ul}$ and illustrate it in Fig.~\ref{Fig:CoordinateSystems}(b). The sinc-function pattern in angular domain is exactly the array pattern of the ULA.

The two properties indicate that the information of $\Gamma_l$, $\Theta_l$, and $\Phi_l$ are directly illustrated in the image of uplink pilots. By observing the dark spots in the image, we can easily obtain these frequency-independent parameters.

\subsection{Power of YOLO network}

With the two properties, we regard the dark spots as the objects and tackle the problems in Section \ref{Sec:YOLO}.B with a powerful neural network for object detection, that is, YOLO.

\emph{Fast}:
As the name suggests, YOLO can find all objects that the network knows in an image by only observing the image once. According to \cite{Redmon2015}, YOLO can process 45 large images in a second, thereby demonstrating its rapid processing ability.

\emph{Ability to bound objects}:
YOLO can position the objects and estimate the size of each object by observing the bounding boxes that frame the objects. If the bounding boxes can be learned to exactly bound the dark spots, then we can answer the two questions as follows.

\begin{itemize}
  \item {\bf Answer to Q1}: $\Gamma_l$ and $\Theta_l$ can be rapidly estimated by calculating the center of the $l$th bounding box.
  \item {\bf Answer to Q2}: The size of $\Phi_l$ can be estimated by observing the height of the $l$th bounding box, thereby simplifying the identification of $\Phi_l$.
\end{itemize}

YOLO has advanced to version 3 \cite{Redmon2018}, which has a comprehensive network structure but a greatly enhanced successful detection ratio of small objects. Therefore, this version is adopted in this study. We maintain the original structure and the input and output settings of the YOLO network to the greatest extent. However, we perform the necessary modifications to satisfy the requirement of parameter estimation.

The image in Fig.~\ref{Fig:CoordinateSystems}(a) illustrates the input of the YOLO network. Only a small amount of data can train the network because all the input images of YOLO have strong similarities. YOLO has its own coordinate system, where the top left vertex of a input image is regarded as the origin (0,0), as shown in Fig.~\ref{Fig:CoordinateSystems}(c). The $x$ and $y$ axes stretch rightward and downward, respectively. These settings coincide with the coordinate system of the image of uplink pilots. Each axis in the coordinate system of the network ranges from 0 to 938,\footnote{938 is the double of the resolution of the network.} and the coordinates take integer values.

Here, the network outputs $5{\hat L}$ parameters after processing the image, where ${\hat L}$ denotes the number of detected paths and is an estimate of $L$. Five parameters are provided to describe the $l$th detected path, which are denoted as
\begin{equation}\label{Eq:YOLOoutput}
\left\{C_{l},x_{l,\min},y_{l,\min},x_{l,\max},y_{l,\max}\right\}
\end{equation}
where $C_{l}$ indicates the confidence level of the detection of the $l$th path, satisfying $0 < C_{l}\le 1$. When $C_{l}$ grows large, the probability of a successful detection increases. Generally, if $C_{l}$ is less than 0.5, then the $l$th detected path may be fake. False alarm generally happens in a low SNR regime, where the noise is falsely identified as the path. Only one class of object (the path) should be recognized; thus, the class indicator in the original network is no longer provided in the output.

Specially, $(x_{l,\min},y_{l,\min})$ and $(x_{l,\max},y_{l,\max})$ are the coordinates of the top left and the bottom right vertexes of the bounding box, respectively, thereby satisfying $0\le x_{l,\min},y_{l,\min} < 938$ and $0<x_{l,\max},y_{l,\max}\le938$, as shown in Fig.~\ref{Fig:CoordinateSystems}(c).
The bounding box exactly bounds the dark spots as suggested. Thereafter, when generating the labels of training data, we set
\begin{equation}\label{Eq:boxCenter}
\begin{aligned}
x_{l,\min} &= \left\lceil 938\left(\Theta_{l}-\frac{w_l}{2}\right)\right\rceil,  y_{l,\min} = \left\lceil 938\left(\Gamma_{l}-\frac{h_l}{2}\right)\right\rceil,\\
x_{l,\max} &= \left\lceil 938\left(\Theta_{l}+\frac{w_l}{2}\right)\right\rceil,  y_{l,\max} = \left\lceil 938\left(\Gamma_{l}+\frac{h_l}{2}\right)\right\rceil.
\end{aligned}
\end{equation}
Under this setting of coordinates, $\Gamma_l$, $\Theta_l$, and $\Phi_l$ can be estimated efficiently.

\subsection{YOLO-based parameter estimation}

Based on the Answer to Q1, $\Theta_l$ and $\Gamma_l$ are derived from the center of the bounding box, i.e.,
\begin{equation}\label{Eq:YOLOestTheta}
{\tilde\Theta}_{l} = \frac{y_{l,\min}+y_{l,\max}}{2\times 938}
\end{equation}
and
\begin{equation}\label{Eq:YOLOestGamma}
{\tilde\Gamma}_{l} = \frac{x_{l,\min}+x_{l,\max}}{2\times 938},
\end{equation}
where $\tilde\Theta_l$ and $\tilde\Gamma_l$ are the coarse estimates of $\Theta_l$ and $\Gamma_l$, respectively.

\begin{figure}
  \centering
  \includegraphics[scale=0.55]{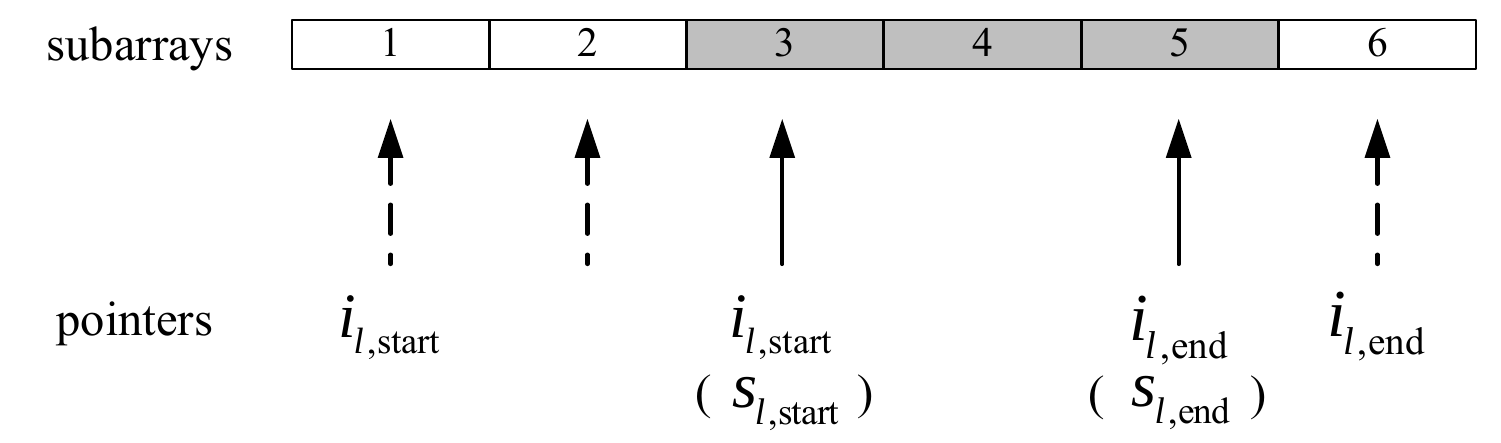}
  \caption{Pointers are used to identify the non-stationarity. The gray blocks represent the subarrays in $\Phi_l$.} \label{Fig:Pointers}
\end{figure}

According to Property \ref{Theo:lightSpot2}, the size of $\Phi_l$ determines the height of the $l$th bounding box, which is calculated as
\begin{equation}\label{Eq:heightYOLOout}
\hat h_l = \frac{y_{l,\max}-y_{l,\min}}{938}.
\end{equation}
For a scatterer that can see $s$ subarrays, the height of the bounding box should be equal to
\begin{equation}\label{Eq:heightHs}
H^{(s)} = \frac{2S}{sM},
\end{equation}
where $s=1,\ldots,S$.
We estimate the size of $\Phi_l$ by exhaustively searching $H^{(1)},\ldots,H^{(S)}$ for the one that has the closest value to $\hat h_l$, i.e.,
\begin{equation}\label{Eq:Sl}
S_l = \arg\min_{s=1,\ldots,S} |\hat h_l-H^{(s)}|.
\end{equation}
The identification of $\Phi_l$ is greatly simplified with the knowledge of $S_l$, because we are required to identify only the first or the last subarrays of the adjacent $S_l$ subarrays. Two pointers, denoted as $i_{l,{\rm start}}$ and $i_{l,{\rm end}}$, are the indicators of $s_{l,{\rm start}}$ and $s_{l,{\rm end}}$, respectively, as shown in Fig.~\ref{Fig:Pointers}. We initially set $i_{l,{\rm start}}=1$ and $i_{l,{\rm end}}=S$. We can determine the $S_l$ subarrays by moving the two pointers for a sum of $S-S_l$ steps.

We decide how to move the pointers by observing the projection power. If a subarray can see a scatterer, then the uplink pilots on this subarray obtain distinct projection power on this path. The projection power from path $l$ to the uplink pilots on subarray $s$ is defined as
\begin{equation}\label{Eq:ProjectPower}
P_{l,s} = \left|\left({\bf a}({\tilde\Theta}_l)\odot{\bf p}(\{s\})\right)^H{\bf Y}^{\rm ul}{\bf q}^*({\tilde\Gamma}_l)\right|^2.
\end{equation}
The following theorem provides the approximation of $P_{l,s}$.

\begin{property}\label{Theo:PsApprox}
If ${\tilde\Theta}_l\approx\Theta_l$, ${\tilde\Gamma}_l\approx\Gamma_l$, and the size of each subarray is large, then $P_{l,s}$ can be approximated by
\begin{equation}\label{Eq:PsApprox}
P_{l,s} \approx
\begin{cases}
P^{\rm ul}|\alpha_l|^2 {M^2N^2}/{S^2} + {MN}/{S}, & \text{if $s \in \Phi_l$,} \\
{MN}/{S}, & \text{else.}
\end{cases}
\end{equation}
\end{property}

\begin{proof}
See Appendix C.
\end{proof}

According to Property \ref{Theo:PsApprox}, $P_{l,s_1}\approx P_{l,s_2}$ holds for subarrays $s_1,s_2\in\Phi_l$. Meanwhile, $P_{l,s_1}\gg P_{l,s_2}$ holds for subarrays $s_1\in\Phi_l$ and $s_2\notin\Phi_l$. That is, the subarrays in $\Phi_l$ have similar values of projection power on path $l$, and these values are much larger than those of the subarrays that are not in $\Phi_l$.

For the two pointers, if $P_{l,i_{l,{\rm start}}} \ge P_{l,i_{l,{\rm end}}}$, then the probability that $i_{l,{\rm end}}\notin\Phi_l$ is high. We move the pointer $i_{l,{\rm end}}$ backward by one step, i.e., $i_{l,{\rm end}} = i_{l,{\rm end}}-1$. Otherwise, we move the pointer $i_{l,{\rm start}}$ forward by one step, i.e., $i_{l,{\rm start}} = i_{l,{\rm start}}+1$. We continue to move the pointers until $i_{l,{\rm end}}-i_{l,{\rm start}}=S_l-1$. Thereafter, we set
\begin{equation}\label{Eq:sstartendEst1}
{\hat s}_{l,{\rm start}}=i_{l,{\rm start}}, {\hat s}_{l,{\rm end}}=i_{l,{\rm end}},
\end{equation}
and obtain the estimate of $\Phi_l$ as follows:
\begin{equation}\label{Eq:PhiEst}
\hat\Phi_l = \{ {\hat s}_{l,{\rm start}}, {\hat s}_{l,{\rm start}}+1, \ldots, {\hat s}_{l,{\rm end}}\}.
\end{equation}
The non-stationarity identification algorithm that utilizes $S_l$ is named as the bounding box-based algorithm.

\begin{figure*}
  \centering
  \includegraphics[scale=0.88]{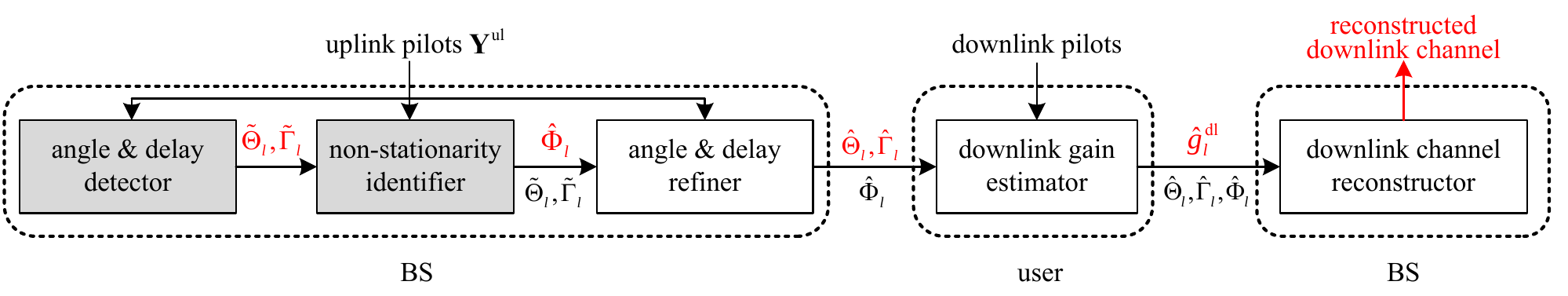}
  \caption{Modules of the proposed downlink channel reconstruction scheme. The modules in gray are based on deep learning. The symbols above the arrow are the outputs of the left module of the arrow.} \label{Fig:SchemeModules}
\end{figure*}

\section{Downlink channel reconstruction scheme}\label{Sec:scheme}

On the basis of the channel model and the power of YOLO, we propose a model-driven deep learning-based scheme to reconstruct the non-stationary downlink channel.
Fig.~\ref{Fig:SchemeModules} illustrates the diagram of the proposed non-stationary downlink channel reconstruction scheme. The scheme functions successively in the following five modules.

Module 1: The \emph{angle and delay detector} at the BS obtains coarse estimates $\tilde\Theta_l$ and $\tilde\Gamma_l$ by applying \eqref{Eq:YOLOestTheta} and \eqref{Eq:YOLOestGamma}.

Module 2: The \emph{non-stationarity identifier} at the BS obtains $\hat\Phi_l$ through the bounding box-based algorithm or the algorithm described in the following section that uses $\tilde\Theta_l$, $\tilde\Gamma_l$, and ${\bf Y}^{\rm ul}$.

Module 3: The \emph{angle and delay refiner} at the BS obtains $\hat\Theta_l$ and $\hat\Gamma_l$, which are the refined estimates of $\Theta_l$ and $\Gamma_l$, respectively, by utilizing $\tilde\Theta_l$, $\tilde\Gamma_l$, $\hat\Phi_l$, and ${\bf Y}^{\rm ul}$.

Module 4: The \emph{downlink gain estimator} at the user obtains ${\hat g}^{\rm dl}_l$, which is the estimate of ${g}^{\rm dl}_l$, by utilizing the downlink pilots and $\hat\Phi_l$, $\hat\Theta_l$, and $\hat\Gamma_l$, and sends ${\hat g}^{\rm dl}_l$ to the BS.

Module 5: The \emph{downlink channel reconstructor} at the BS reconstructs the downlink channel by applying $\hat\Phi_l$, $\hat\Theta_l$, $\hat\Gamma_l$, and ${\hat g}^{\rm dl}_l$ to \eqref{Eq:DLchannel} as follows:
\begin{equation}\label{Eq:DLchannelrec}
{\hat{\bf H}}^{\rm dl} = \sum_{l=1}^{\hat L} {\hat g}^{\rm dl}_l \left({\bf a}(\hat\Theta_l) \odot {\bf p}(\hat\Phi_l)\right) {\bf q}^T(\hat\Gamma_l).
\end{equation}

The proposed scheme has low overhead and low complexity. In comparison with \cite{Han2019TWC}, the present work can rapidly identify the non-stationarity aside from detecting delays and angles. In the following subsections, modules 2--4 are described in detail, and the reduced case in the stationary scenario is further discussed.

\subsection{Non-stationarity identifier}

The bounding box-based algorithm utilizes the deep learning results but is sensitive to the accuracy of $y_{l,\max}$ and $y_{l,\min}$, especially when the size of $\Phi_l$ is smaller than $S$ and the power of this path is much smaller than the largest power of a path in the channel (i.e., $|\alpha_l|\ll\max_k|\alpha_k|$). To enhance the accuracy of the non-stationarity identifier, we further propose a projection power-based algorithm.

This algorithm is also based on Property \ref{Theo:PsApprox}, but identifies $\Phi_l$ by comparing the projection power from path $l$ to the uplink pilots on each subarray. We initially determine the subarray with the maximal projection power on path $l$,
\begin{equation}\label{Eq:barsl}
\bar s_l = \arg\max_{s=1,\ldots,S} P_{l,s}.
\end{equation}
Subsequently, we find the subarrays that have similar projection power with $P_{l,\bar s_l}$.

We still introduce two pointers and initialize them by $j_{l,{\rm start}}=1$ and $j_{l,{\rm end}}=S$. We move forward the pointer $j_{l,{\rm start}}$ until $P_{l,j_{l,{\rm start}}} \ge \delta P_{l,\bar s_l}$, where $0<\delta<1$. We set $\delta \in [0.1,0.5]$, considering the estimation error of ${\tilde\Theta}_l$ and ${\tilde\Gamma}_l$ and the existence of noise. Afterward, we move backward the pointer $j_{l,{\rm end}}$ until $P_{l,j_{l,{\rm end}}} \ge \delta P_{l,\bar s_l}$. Finally, the estimated indices of the first and last subarrays in $\Phi_l$ are
\begin{equation}\label{Eq:sstartendEst2}
{\hat s}_{l,{\rm start}}=j_{l,{\rm start}}, {\hat s}_{l,{\rm end}}=j_{l,{\rm end}},
\end{equation}
and $\hat\Phi_l$ is derived by applying \eqref{Eq:sstartendEst2} to \eqref{Eq:PhiEst}.

\subsection{Angle and delay refiner}

With ${\tilde\Theta}_l$, ${\tilde\Gamma}_l$, and $\hat\Phi_l$, the angle and delay refiner then calculates the refined estimates of angles and delays. Prior to describing the method to refine the estimates, we initially explain the reason of introducing this module.

\subsubsection{Reasons of introducing the refiner}

The angle and delay refiner is introduced because the accuracy of ${\tilde\Theta}_l$ and ${\tilde\Gamma}_l$ is impacted by the following factors of YOLO.

\emph{Image resolution}:
Each image is generated by a finite-dimensional angular temporal domain pilot matrix. The values of $\gamma_{\rm a}M$ and $\gamma_{\rm t}N$ are large but not infinite, thereby resulting in the on-grid effect. Then, the coordinates of the $l$th dark spot center are close to but not equal to $(\Theta_l,\Gamma_l)$. One solution is to increase the values of $\gamma_{\rm a}$ and $\gamma_{\rm t}$. However, scaling the oversampling rates results in multiplied complexity and extended running time to generate the images.

\emph{Network resolution}:
The maximal coordinates in the coordinate system of network are (938,938). For any input image, the network initially rescales the size of the image to $938\times 938$. That is, a $938\times 938$ dimensional matrix is processed in the network, instead of the original $\gamma_{\rm a}M\times\gamma_{\rm t}N$ dimensional matrix. Once $938<\gamma_{\rm a}M$ or $938<\gamma_{\rm t}N$ holds, the resolution is decreased.

\emph{Integer labels}:
We set the coordinates of the bounding boxes as integers to maintain the settings of the original YOLO network and guarantee the accuracy of detection. Using integer coordinates also results in the on-grid effect.

\emph{Detection error}:
Although the network is well trained, the detection error is inevitable. The bounding box may deviate from the ideal one. The minimum deviation amount is 1, thereby resulting in the error amount of ${1}/{938}$. Moreover, false alarm and miss detection may occur in a low SNR regime. Therefore, the network detection error is the most critical factor that harms the accuracy.

Consequently, the accuracy of ${\tilde\Theta}_l$ and ${\tilde\Gamma}_l$ is questioned due to these factors. Especially when $M$ and $N$ are large, a small error of angle and delay results in sharp degradation of channel reconstruction accuracy. Therefore, further processing these coarse estimates is necessary.

\subsubsection{Refining the estimates}

The inputs of the angle and delay refiner are
\begin{equation}\label{Eq:RefineInput}
\left\{{\bf Y}^{\rm ul}, {\tilde\Theta}_1, {\tilde\Gamma}_1, {\hat\Phi}_1, \ldots, {\tilde\Theta}_{\hat L}, {\tilde\Gamma}_{\hat L}, {\hat\Phi}_{\hat L} \right\}
\end{equation}
The outputs are the refined angles and delays, as follows:
\begin{equation}\label{Eq:RefineOutput}
\left\{{\hat\Theta}_1, {\hat\Gamma}_1, \ldots, {\hat\Theta}_{\hat L}, {\hat\Gamma}_{\hat L} \right\}.
\end{equation}
Recalling the NOMP algorithm, within each iteration, NOMP refines all the extracted paths through the Newton refinement method, which can effectively refine the estimates of delays and angles toward their real values. However, the original Newton refinement method is designed for stationary systems. Here, we adjust the method to fit the non-stationary cases.

The Newton method refines the paths one by one in decreasing order of the path power to guarantee the effectiveness of refinement. We initially calculate the coarse estimates of uplink effective gains of these $\hat L$ paths by
\begin{equation}\label{Eq:YOLOestULGain}
\left[{\tilde\alpha}_1, \ldots, {\tilde\alpha}_{\hat L} \right]^T = \left( {\bf A}^{{\rm ul}H} {\bf A}^{\rm ul} \right)^{-1} {\bf A}^{{\rm ul}H} {\bf y}^{\rm ul},
\end{equation}
where ${\bf A}^{\rm ul}\in\mathbb{C}^{MN\times \hat L}$, the $l$th column of ${\bf A}^{\rm ul}$ is
\begin{equation}\label{Eq:YOLOestULmtxA}
[{\bf A}^{\rm ul}]_{:,l} = {\bf q}({\tilde\Gamma}_{l})\otimes \left({\bf a}({\tilde\Theta}_{l})\odot {\bf p}({\hat\Phi}_{l})\right),
\end{equation}
and ${\bf y}^{\rm ul}$ is obtained by stacking all the columns of ${\bf Y}^{\rm ul}$ into a vector. Thereafter, we sort these paths by the decreasing order of $\|{\tilde\alpha}_l {\bf p}(\hat\Phi_l)\|^2$. To simplify the expression, we still maintain the denotations of the coarse estimates in \eqref{Eq:RefineInput}, which currently satisfy $\|{\tilde\alpha}_1 {\bf p}(\hat\Phi_1)\|^2 \ge \ldots \ge \|{\tilde\alpha}_{\hat L} {\bf p}(\hat\Phi_{\hat L})\|^2$. The residue is calculated as
\begin{equation}\label{Eq:ULpilotsRes}
{\bf Y}^{\rm ul}_{{\rm res}} = {\bf Y}^{\rm ul} - \sum_{l=1}^{\hat L} \sqrt{P^{\rm ul}}{\tilde\alpha}_l \left({\bf a}({\tilde\Theta}_{l})\odot {\bf p} ({\hat\Phi}_{l})\right)  {\bf q}^T({\tilde\Gamma}_{l}).
\end{equation}
The Newton method refines the angles and delays by minimizing the residue power.

We describe the Newton method by taking the first path as an example. We initially define
\begin{equation}\label{Eq:ULpilotsNT1}
{\bf Y}^{\rm ul}_{{\rm res},+1} = {\bf Y}^{\rm ul}_{{\rm res}} + \sqrt{P^{\rm ul}} {\tilde\alpha}_1 \left({\bf a}({\tilde\Theta}_{1})\odot {\bf p} ({\hat\Phi}_{1})\right)  {\bf q}^T({\tilde\Gamma}_{1}).
\end{equation}
Only the uplink pilots on the subarrays in ${\hat\Phi}_{1}$ are utilized in the refinement of ${\tilde\Theta}_1$ and ${\tilde\Gamma}_1$. The refined estimates obtained by the Newton method, that is, ${\hat{\alpha}}_1$, ${\hat{\Theta}}_1$, and ${\hat{\Gamma}}_1$, can achieve the minimum residue power, i.e.,
\begin{equation}\label{Eq:ULpilotsNT2}
\begin{aligned}
&({\hat{\alpha}}_1, {\hat{\Theta}}_1, {\hat{\Gamma}}_1) \\= &\arg\min_{\alpha,\Theta,\Gamma} \left\|\left[{\bf Y}^{\rm ul}_{{\rm res},+1} - \sqrt{P^{\rm ul}} {\alpha}{\bf a}({\Theta}){\bf q}^T({\Gamma})\right]_{{\bf r}({\hat\Phi}_{1}),:} \right\|^2_F,
\end{aligned}
\end{equation}
where the row-selection vector is defined as
\begin{equation}\label{Eq:rPhi}
{\bf r}(\Phi) = \left[ \frac{M}{S}(s_{\rm start}-1)+1,\ldots, \frac{M}{S}s_{\rm end}\right],
\end{equation}
and $s_{\rm start}$ and $s_{\rm end}$ represent the indices of the first and last subarrays in $\Phi$, respectively.
The derivations of ${\hat{\alpha}}_1$, ${\hat{\Theta}}_1$, and ${\hat{\Gamma}}_1$ are the same as the original Newton refinement method in \cite{Han2019TWC}; thus, they are omitted here. Having refined the estimates of the first path, we update the residue by
\begin{equation}\label{Eq:ResUpdate}
{\bf Y}^{\rm ul}_{{\rm res}} = {\bf Y}^{\rm ul}_{{\rm res},+1} - \sqrt{P^{\rm ul}}{\hat\alpha}_1 \left({\bf a}({\hat\Theta}_{1})\odot {\bf p} ({\hat\Phi}_{1})\right)  {\bf q}^T({\hat\Gamma}_{1}).
\end{equation}
Thereafter, the estimates of the other paths are refined following the similar approach from \eqref{Eq:ULpilotsNT1}. The angle and delay refiner repeats the above refinement methods for $R_c$ rounds. The refinement has a low complexity of $\mathcal{O}(R_c \hat LMN)$.

After the refiner completes its work, all the frequency-independent parameters are acquired by the BS. The BS then reconstructs the uplink channel by
\begin{equation}\label{Eq:ULrecChannel}
{\hat{\bf H}}^{\rm ul} = \sum_{l=1}^{\hat L} {\hat\alpha}_l \left({\bf a}({\hat\Theta}_{l})\odot {\bf p} ({\hat\Phi}_{l})\right)  {\bf q}^T({\hat\Gamma}_{l}).
\end{equation}

\subsection{Downlink gain estimator}

The estimated frequency-independent parameters, including $\hat\Theta_l$, $\hat\Gamma_l$, and $\hat\Phi_l$, are sent to the downlink gain estimator. This module functions at the user equipment. As suggested in \cite{Han2019TWC}, the downlink pilots are beamformed along the angles of the paths to enhance the received power at user equipment and improve the estimation accuracy of downlink gains. Thus, all-one downlink pilots occupy $\hat L$ OFDM symbols. The pilots received by the user on OFDM symbol $t$ are expressed as
\begin{equation}\label{Eq:DLpilots}
{\bf y}^{\rm dl}_{t} = \sum_{l=1}^{L} \sqrt{P^{\rm dl}} g^{\rm dl}_{l} {\bf q}(\Gamma_{l}) \left({\bf a}(\Theta_l)\odot{\bf p}(\Phi_l)\right)^T  {\bf b}_t +{\bf z}^{\rm dl}_{t},
\end{equation}
where $t = 1,\ldots,\hat L$, ${\bf y}^{\rm dl}_{t}\in\mathbb{C}^{N\times 1}$, $P^{\rm dl}$ is the transmitted power of BS,
\begin{equation}\label{Eq:DLpilotsBF}
{\bf b}_t = \sqrt{\frac{S}{(\hat s_{t,{\rm end}}-\hat s_{t,{\rm start}}+1)M}}{\bf a}^*(\hat\Theta_t)\odot{\bf p}^T(\hat\Phi_t)
\end{equation}
is the beamforming vector for the downlink pilots on OFDM symbol $t$, and ${\bf z}^{\rm dl}_{t}\in\mathbb{C}^{N\times 1}$ is the downlink complex Gaussian noise whose elements are i.i.d. with zero mean and unit variance. $\|{\bf b}_t\|^2=1$ due to the power constrain. The design in \eqref{Eq:DLpilotsBF} indicates that on OFDM symbol $t$, the transmitted power is allocated only to the subarrays in $\hat\Phi_t$.

Afterwards, $\hat\Theta_l$, $\hat\Gamma_l$, and $\hat\Phi_l$ are applied in \eqref{Eq:DLpilots} to replace $\Theta_l$, $\Gamma_l$, and $\Phi_l$, respectively. The downlink gains of the $\hat L$ paths are estimated by
\begin{equation}\label{Eq:estDLGain}
\left[{\hat g}^{\rm dl}_1, \ldots, {\hat g}^{\rm dl}_{\hat L} \right]^T = \left( {\bf A}^{{\rm dl}H} {\bf A}^{\rm dl} \right)^{-1} {\bf A}^{{\rm dl}H} {\bf y}^{\rm dl},
\end{equation}
where
\begin{equation}\label{Eq:estDLmtxA}
{\bf A}^{\rm dl} = \left[ {\bf A}^{{\rm dl}T}_1,\ldots,{\bf A}^{{\rm dl}T}_{\hat L} \right]^T,
\end{equation}
the $l$th column of the $t$th submatrix ${\bf A}^{\rm dl}_t \in \mathbb{C}^{N\times \hat L}$ is expressed as
\begin{equation}\label{Eq:estDLmtxAt}
[{\bf A}^{\rm dl}_t]_{:,l} = {\bf q}(\hat\Gamma_l) \left({\bf a}(\hat\Theta_l)\odot{\bf p}(\hat\Phi_l)\right)^T  {\bf b}_t,
\end{equation}
and ${\bf y}^{\rm dl}$ is obtained by stacking ${\bf y}^{\rm dl}_{1},\ldots,{\bf y}^{\rm dl}_{\hat L}$ into a vector.

The estimated downlink gains are fed back to the BS. Finally, the proposed scheme is completed with ${\hat{\bf H}}^{\rm dl}$ being reconstructed by the downlink channel reconstructor at the BS.

\begin{figure*}
  \centering
  \includegraphics[scale=0.82]{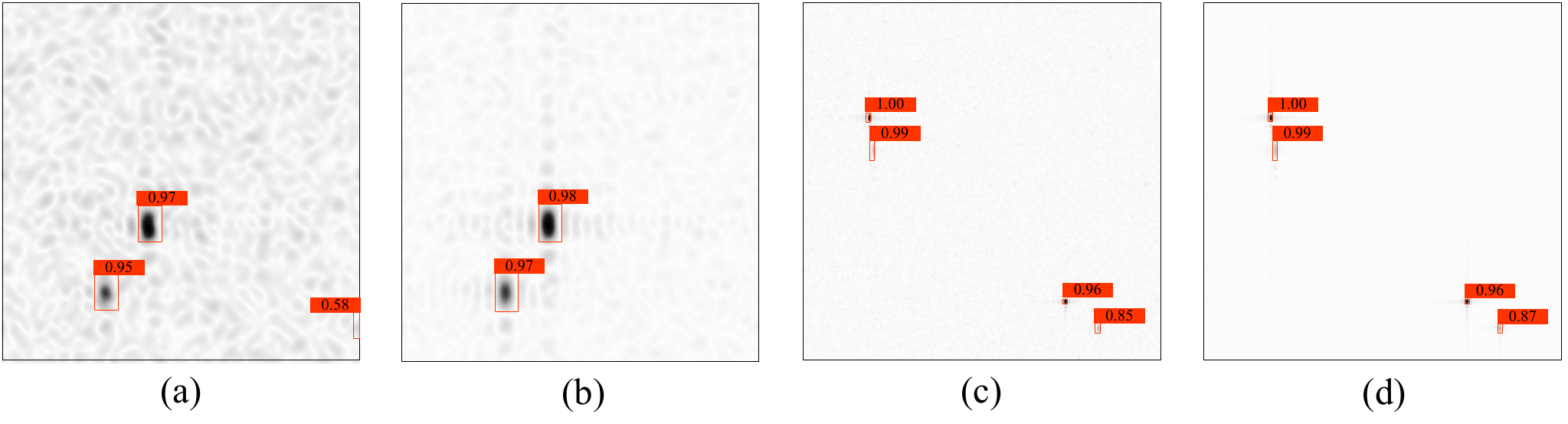}
  \caption{YOLO detection results of the images of uplink pilots when $S=4$, (a) $M=N=32$, SNR$=0$dB, (b) $M=N=32$, SNR$=10$dB, (c) $M=N=128$, SNR$=0$dB, and (d) $M=N=128$, SNR$=10$dB. The values upon the bounding boxes are the confidences of detection.} \label{Fig:YOLOtest}
\end{figure*}

\subsection{Discussions}

The proposed scheme can efficiently reconstruct the FDD non-stationary massive MIMO downlink channel by identifying the mapping between scatterers and subarrays. Moreover, it can be reduced to fit the stationary systems, thereby further achieving an alternative scheme to reconstruct the non-stationary downlink channel.

\subsubsection{Reducing to stationary cases}

In stationary massive MIMO systems, each scatterer can see all the subarrays. Or equivalently, the ULA is segmented into only $S=1$ subarray, and $\Phi_1=\ldots=\Phi_{L}=\{1\}$ holds. Under this condition, the following changes occur for the proposed scheme.

First, in the image of uplink pilots, when $M$ and $N$ are fixed, all the dark spots have the same shape and the unified height or width. With equal size of objects, the bounding boxes can frame the dark spots more accurately.

Second, the outputs of the non-stationarity identifier become $\hat\Phi_1=\ldots=\hat\Phi_{\hat L}=\{1\}$ with probability 1.

Third, the angle and delay refiner and the downlink gain estimator are reduced to the versions for stationary systems.

Therefore, in widely concerned FDD stationary massive MIMO systems, the proposed downlink channel reconstruction scheme also works, and even works better than in non-stationary systems with the same array scale.

\subsubsection{Alternative scheme for non-stationary systems}

The applicability of the proposed scheme in stationary systems inspires us with an alternative scheme to reconstruct the downlink non-stationary channel. It is known that a subarray is the smallest unit to describe the non-stationarity. If one subarray is considered individually, then stationarity exists in the subsystem formed by this subarray. We can apply the proposed scheme in each subsystem individually. Under this condition, the following changes should be performed.

First, the uplink pilots are divided by ${\bf Y}^{\rm ul} = \left[ {\bf Y}^{{\rm ul}T}_1,\ldots,{\bf Y}^{{\rm ul}T}_S \right]^T$, where ${\bf Y}^{{\rm ul}}_s \in\mathbb{C}^{M/S\times N}$.

Second, for subsystem $s$, the image is generated from ${\bf Y}^{{\rm ul}}_s$. The refined angles and delays of the paths that subarray $s$ can see are $\hat\Theta_{s,l}$ and $\hat\Gamma_{s,l}$, respectively, where $l=1,\ldots,\hat L_s$, and $\hat L_s$ is an estimate of $L_s$.

Third, a total of $\sum_{s=1}^{S}\hat L_s$ paths are estimated by the alternative scheme, requiring $\sum_{s=1}^{S}\hat L_s$ OFDM symbols for downlink pilots. Afterwards, the downlink gains, denoted as $\hat g^{\rm dl}_{s,l}$, are sent back to the BS, where $l=1,\ldots,\hat L_s, s=1,\ldots,S$.

Fourth, the stationary downlink channel in subsystem $s$ is reconstructed using $\hat\Theta_{s,l}$, $\hat\Gamma_{s,l}$, and $\hat g^{\rm dl}_{s,l}$, where $l=1,\ldots,\hat L_s$. Thereafter, the large-scale non-stationary channel is obtained by stacking all the stationary downlink channels together.

The alternative scheme requires a large amount of downlink training and feedback overhead because $\sum_{s=1}^{S}\hat L_s>L$. Moreover, when $\Psi_1=\Psi_2$ holds, the alternative scheme cannot identify this equivalence. The estimation accuracy of angles and delays further degrades when using a reduced number of antennas. Therefore, the proposed scheme is more efficient than the alternative.

\section{Numerical results}\label{Sec:simulations}

In this section, we evaluate the performance of the proposed non-stationary downlink channel reconstruction scheme. In the FDD system, $f^{\rm ul}=2.58$ GHz, and $f^{\rm dl}=2.64$ GHz. The OFDM subcarrier spacing is $\Delta f=15$ kHz. The number of paths $L$ is uniformly distributed in $[1,10]$. For the $l$th path, $\Theta_l$ and $\Gamma_l$ are uniformly distributed in $[0,1)$. The effective uplink gain satisfies $\alpha_l = \beta_l e^{j\phi^{\rm ul}_l}$, where $\beta_l$ is uniformly distributed in $[0.5,1]$, and $\phi^{\rm ul}_l$ is uniformly distributed in $[0,2\pi)$. The downlink gain is $g^{\rm dl}_l = \beta_l e^{j\phi^{\rm dl}_l}$, where $\phi^{\rm dl}_l$ is i.i.d. with $\phi^{\rm ul}_l$.

YOLO is implemented on the computer with one Nvidia GeForce GTX 1080 Ti GPU. The deep learning library of Keras running on top of TensorFlow is used. In the training phase, we generate 3,000 groups of data under each set of $M$, $N$, and $S$. Each group of training data consists of an image of uplink pilots and a label vector, which is denoted as
\begin{equation}\label{Eq:YOLOlabel}
\left\{0, x_{l,\min},y_{l,\min},x_{l,\max},y_{l,\max} \right\},
\end{equation}
where the first parameter (0) indicates the object class. We set $\gamma_{\rm a}=\gamma_{\rm t}=16$, and $\eta=255$. $L$ is randomly and uniformly distributed in $[1,10]$. SNR ranges from 0 dB to 10dB. The number of epochs is 300, and the batch size is 4. The training and testing data are generated by following the same procedure described in Section \ref{Sec:YOLO} and are not biased from each other. Therefore, overfitting issues do not exist.

\subsection{Evaluation of deep learning-based estimation}

We initially test the performance of the angle and delay detector, especially the detection accuracy of the YOLO network. For any input image, the ratio of successful detection of objects is increased when the sizes of objects are large. Thus, we start from the large dark spot cases, where $M=N=32$ and $S=4$. The channel is composed of two paths, satisfying $\Theta_1=0.6195$, $\Gamma_1=0.4102$, and $\Phi_1=\{3,4\}$ for path 1 and $\Theta_2=0.8099$, $\Gamma_2=0.2909$, and $\Phi_2=\{1,2\}$ for path 2. Fig.~\ref{Fig:YOLOtest}(a) illustrates the detection result under the condition of SNR $=0$ dB. The figure shows that the network can successfully recognize the actual dark spots from the noisy image with confidence levels of 0.97 and 0.95. The coarse estimates of angles and delays are $\tilde\Theta_1=0.5840$, $\tilde\Gamma_1=0.3865$, and $\hat\Theta_2=0.7625$, $\tilde\Gamma_2=0.2735$, which are close to the actual values. However, in a low SNR regime, the noise is distinct in the image and appears to be similar to the dark spots, thereby resulting in a false alarm with a confidence level of 0.58. Fig.~\ref{Fig:YOLOtest}(b) shows the detection result when the SNR is increased to 10 dB. The cross-style patterns can be clearly observed from the image, and the network detects the dark spots with confidence levels of 0.98 and 0.97. The confidences increase in proportion to SNR, and a false alarm is avoided. However, the coarse estimates of angles and delays are $\tilde\Theta_1=0.5835$, $\tilde\Gamma_1=0.3860$, and $\tilde\Theta_2=0.7625$, $\tilde\Gamma_2=0.2720$, whose accuracy is not improved accordingly.

\begin{figure}
  \centering
  \includegraphics[scale=0.55]{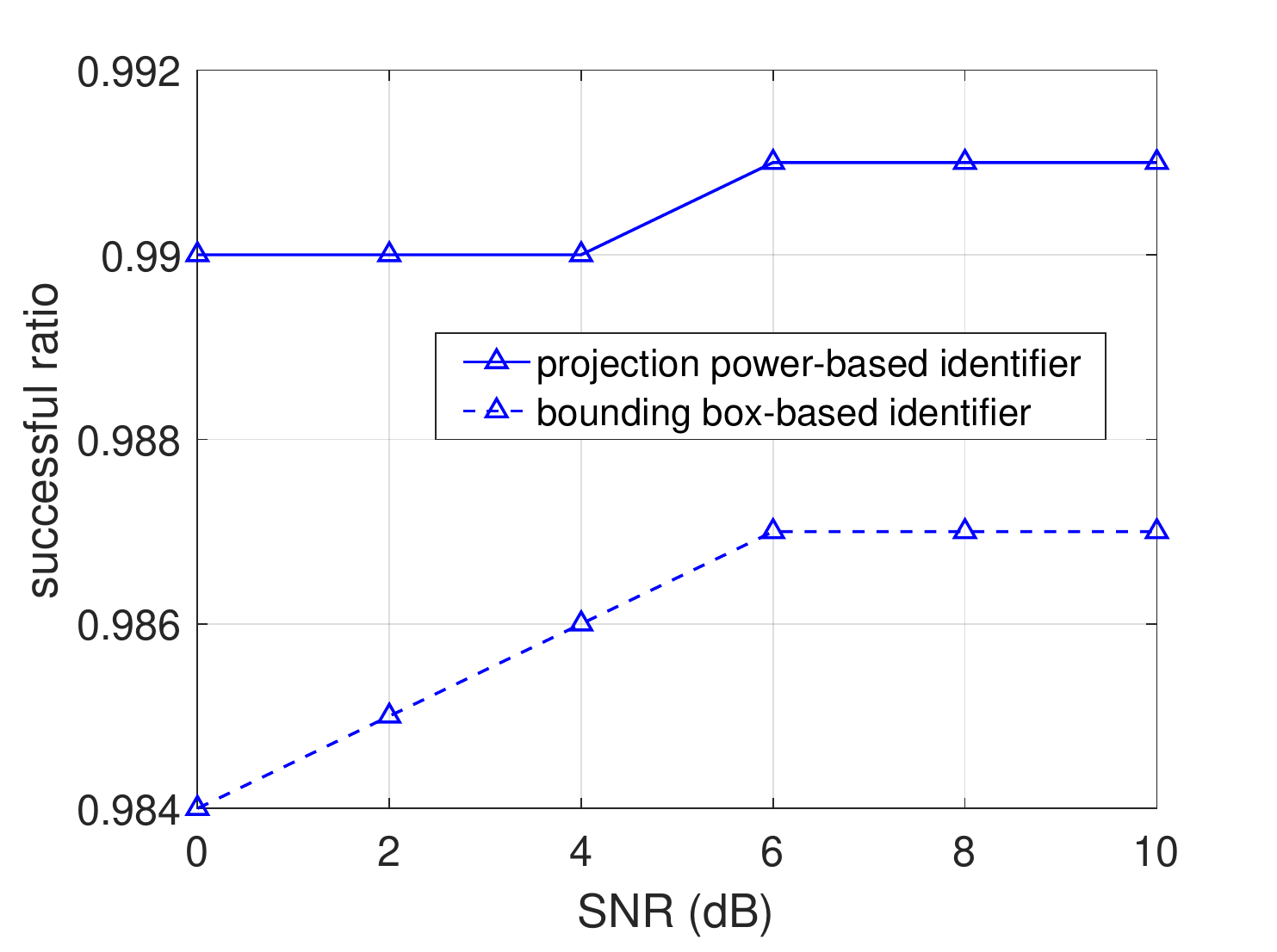}
  \caption{Successful ratios of the algorithms in a non-stationarity identifier.} \label{Fig:IdentifierSuccess}
\end{figure}

Thereafter, the detection accuracy is tested under small dark spot condition, which is the actual massive MIMO condition. We set $M=N=128$ and $S=4$. Figs.~\ref{Fig:YOLOtest}(c) and (d) illustrate the results of SNR $=0$ and $10$ dB, respectively. In massive MIMO systems, the channel becomes sparse, and the noise power is no longer comparable with that of the dark spots even in a low SNR regime. The images of uplink pilots appear to be same as those under SNR $=0$ and $10$ dB. Thus, Figs.~\ref{Fig:YOLOtest}(c) and (d) show similar detection results. The sizes of the bounding boxes are much smaller than those in Figs.~\ref{Fig:YOLOtest}(a) and (b), achieving accurate coarse estimates of angles and delays. We take the first path with a confidence of 1.00 as an example. The actual values are $\Theta_1=0.3229$ and $\Gamma_1=0.1848$. The coarse estimates are $\tilde\Theta_1=0.3015$ and $\tilde\Gamma_1=0.1730$ under SNR$=0$ dB. The accuracy is enhanced compared with that of $M=N=32$. However, the large-aperture array is sensitive to the error of angles. Thus, the refinement of angles and delays is essential.

\begin{figure}
  \centering
  \includegraphics[scale=0.55]{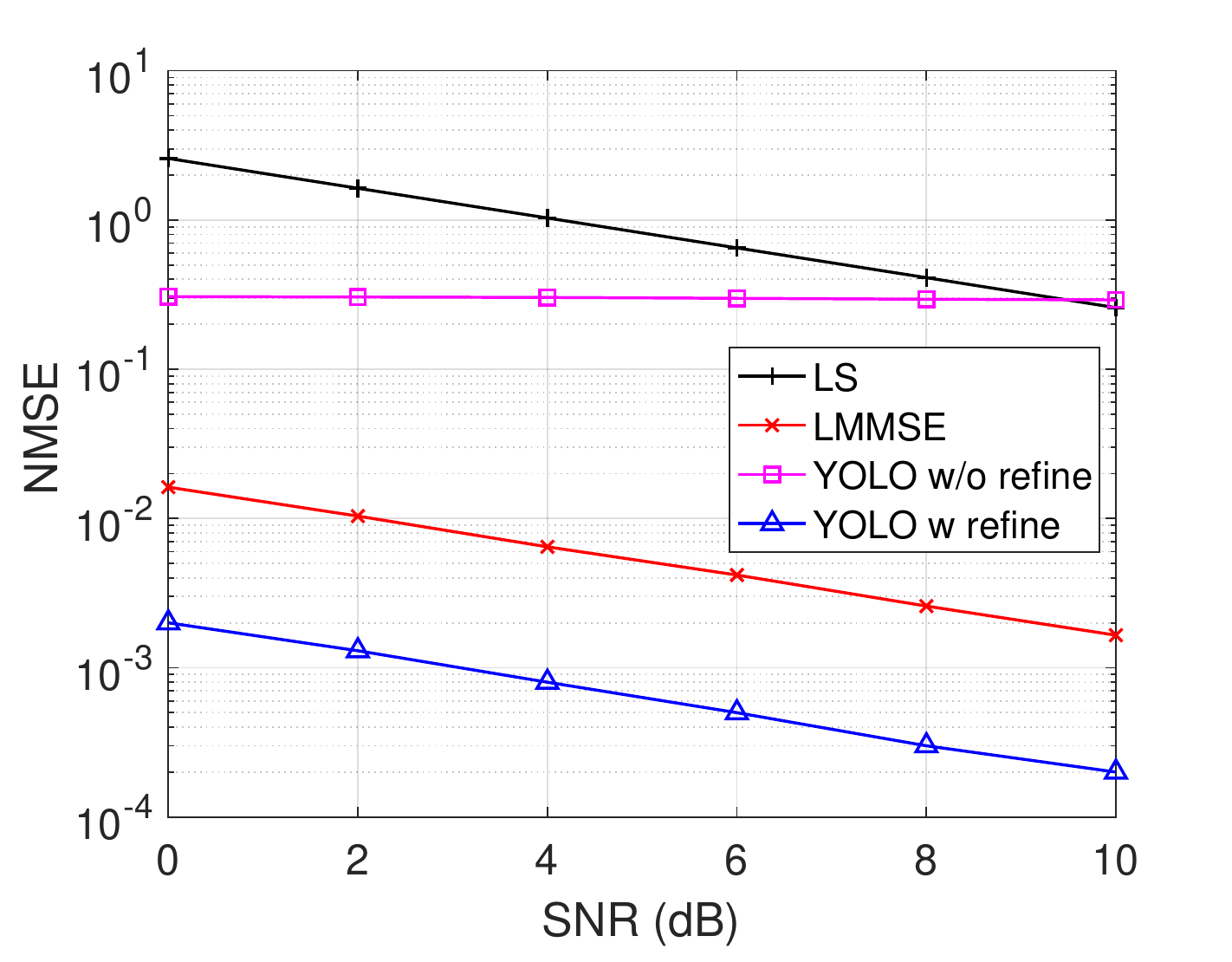}
  \caption{Effectiveness of the angle and delay refiner.} \label{Fig:ULMSE}
\end{figure}

We evaluate the non-stationarity identifier by examining the successful ratio of visibility region identification. For path $l$, the visibility region is successfully identified if $\hat\Phi_l=\Phi_l$. The successful identification ratios of the bounding box-based and the projection power-based algorithms are illustrated in Fig.~\ref{Fig:IdentifierSuccess}, where $M=N=128$, $S=4$, $L=10$, and $\delta=0.2$. The two algorithms can successfully identify the visibility regions with a probability higher than 0.98. As expected, the accuracy of bounding box-based algorithm is sensitive to the detection errors of bounding boxes, whereas the projection power-based algorithm is more robust. Therefore, the projection power-based algorithm achieves a high successful identification ratio. In the following simulations, we adopt this algorithm in the non-stationarity identifier.

\subsection{Evaluation of the refinement}

We examine the effectiveness of the angle and delay refiner through testing the NMSE performance of the reconstructed uplink channel. The refiner works for $R_c=3$ rounds. We introduce two widely used channel estimation algorithms, i.e., LS and LMMSE, as the benchmarks. Notably, LS and LMMSE are not realized through deep learning and do not involve network training. The non-stationary massive MIMO is still considered, where $M=N=128$, and $S=4$. The NMSE is calculated by averaging the NMSEs of the reconstructed or estimated uplink channel across all antennas and on one subcarrier as
\begin{equation}\label{Eq:NMSE}
{\rm NMSE} = \mathbb{E}\left\{\frac{1}{N} \sum_{n=1}^N\frac{\| [\hat{\bf H}^{\rm ul}]_{:,n}-[{\bf H}^{\rm ul}]_{:,n} \|^2}{\| [{\bf H}^{\rm ul}]_{:,n} \|^2} \right\}.
\end{equation}
Fig.~\ref{Fig:ULMSE} illustrates the results. In non-stationary systems, the LS algorithm performs worse than in stationary systems because some subarrays may not see any path, and the channel across these subarrays is zero. However, the LS algorithm still results in a nonzero estimated channel, which is the noise. The LMMSE algorithm identifies the noise through multiplying the covariance matrix of the channel. Thus, the LMMSE algorithm has accurate channel estimation results even in non-stationary systems. For the proposed reconstruction scheme, if we directly apply the coarse estimates in the uplink channel model, then the NMSE of the reconstructed channel is worse. Fig.~\ref{Fig:YOLOtest} shows that even though the coarse estimates of the angles and delays are very close to their actual values, their estimation errors are large and unacceptable in massive MIMO systems, where a small angle offset dramatically impacts the channel reconstruction accuracy. Moreover, the accuracy of the reconstruction without refinement remains $10^{-0.8}$ (i.e., $-8$ dB) and cannot be improved with the increase of SNR because the bounding boxes remain unchanged [Figs.~\ref{Fig:YOLOtest}(c) and (d)]. Fortunately, the angle and delay refiner significantly improves the NMSE of the reconstructed uplink channel, for example, $10^{-2.8}$ (i.e., $-28$ dB) at SNR $=0$ dB. Moreover, the NMSE can be further decreased with the increase in SNR, demonstrating the effectiveness of the refiner.

\begin{figure}
  \centering
  \includegraphics[scale=0.54]{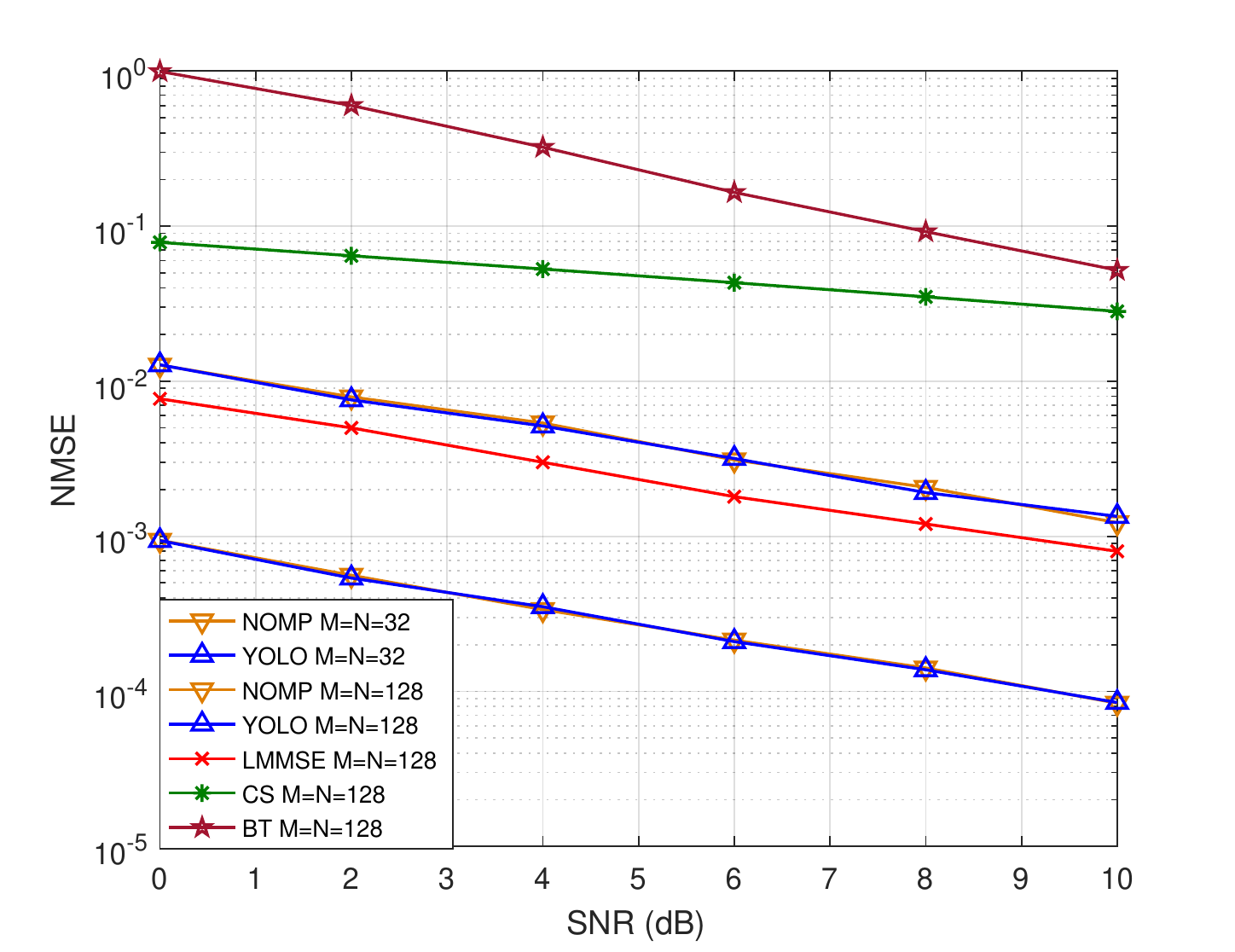}
  \caption{NMSEs of downlink channel estimation schemes under stationarity.} \label{Fig:YOLOvsNOMP}
\end{figure}

\subsection{Comparison of the proposed and the alternative schemes}

We further evaluate the proposed downlink channel reconstruction scheme when reduced to stationary conditions. The NOMP, LMMSE, beam tracking (BT), and compressed sensing (CS)-based downlink channel estimation schemes are introduced as benchmarks. The three latter schemes do not utilize spatial reciprocity and simply rely on downlink training and feedback \cite{Rao2014,Gao2015}. Here, we consider the upper bound cases of the BT and CS-based schemes. The BT-based scheme adopts full-set discrete Fourier transform beams at the BS, and therefore requires $M$ orthogonal downlink pilots. Subsequently, the user feeds back the received pilots that occupy more than 99\% of the total received power and their beam indices. The CS-based scheme also uses $M$ orthogonal downlink pilots to distinguish different BS antennas and adopts the OMP algorithm to estimate the downlink gains on the extracted orthogonal paths, which are then fed back to the BS. The feedback amounts of the two schemes are definitely larger than that of the proposed scheme because of the on-grid effect. In the stationary system, $S=1$, $L\in[1,10]$, and we set $M=N=32$ and $M=N=128$, respectively. Fig.~\ref{Fig:YOLOvsNOMP} presents the NMSEs of the reconstructed or estimated downlink channels. Under the same system settings, the proposed scheme achieves nearly the same accuracy as the NOMP-based scheme with greatly reduced time consumption. Especially when $M=N=128$ and $L=10$, NOMP consumes more than 5 minutes and YOLO costs less than 2 seconds to determine all the paths. Moreover, although with much lower cost of downlink training and feedback, the proposed scheme still significantly outperforms the LMMSE, BT, and CS-based schemes. Therefore, the proposed deep-learning scheme is more efficient than the existing algorithm-based schemes. On the other hand, the accuracy is improved proportional to the values of $M$ and $N$. When SNR $=10$ dB, the NMSE approximates $10^{-3}$ and $10^{-4}$ under the conditions of $M=N=32$ and $M=N=128$, respectively, serving as the lower and upper NMSE bounds of the proposed non-stationary channel reconstruction scheme.

\begin{figure}
  \centering
  \includegraphics[scale=0.55]{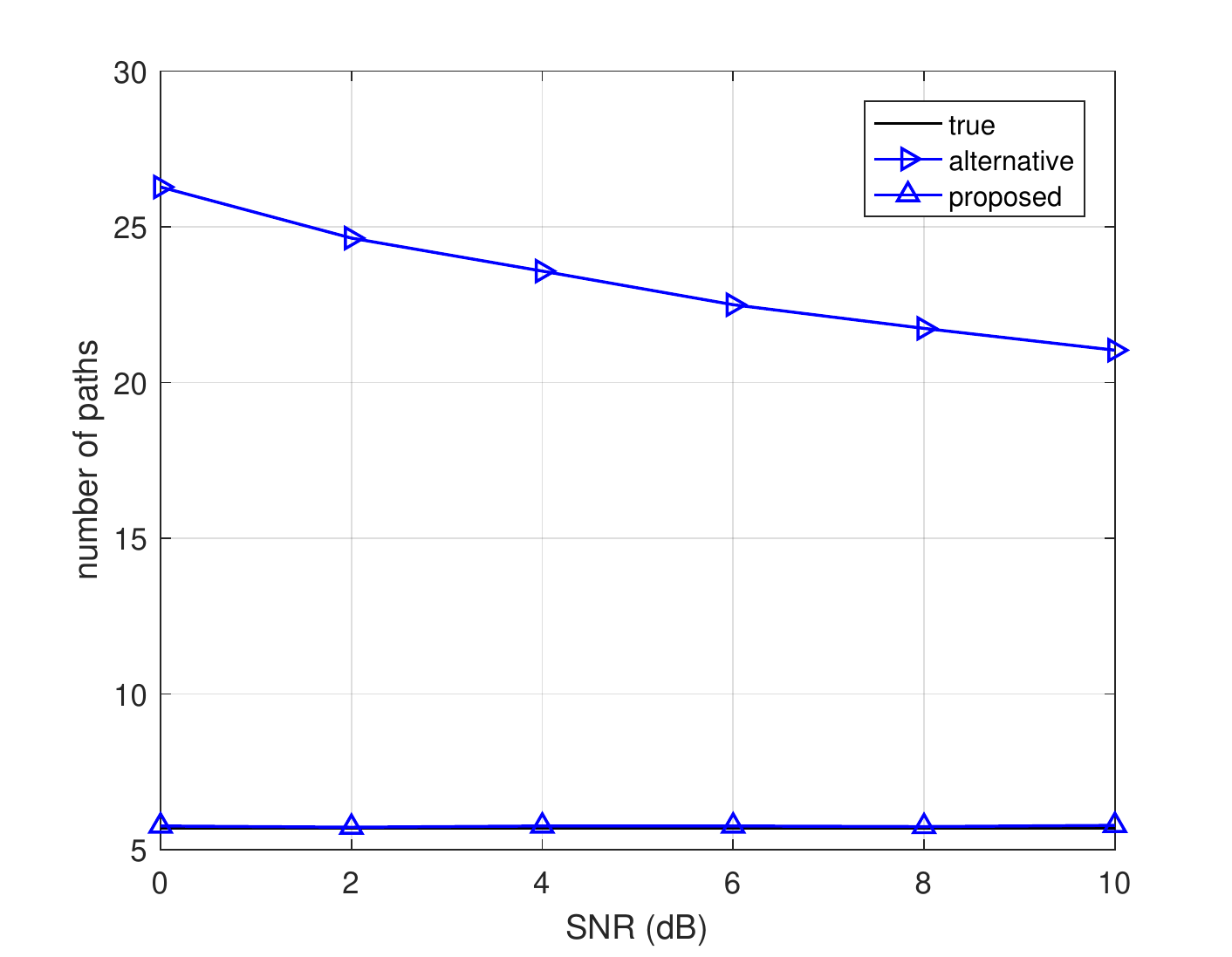}
  \caption{Number of paths estimated by the proposed and alternative schemes.} \label{Fig:SchemesL}
\end{figure}

Finally, we compare the proposed scheme with the alternative scheme under non-stationary conditions of $S=4$, $M=N=128$, and $L\in[1,10]$. From the image drawn by ${\tilde{\bf Y}}^{\rm ul}$, YOLO can detect all the paths without causing missing or false alarm, demonstrating the accuracy of the proposed scheme. The total number of paths estimated by the alternative scheme is more than the number of actual paths and that of the paths estimated by the proposed scheme even though all the paths are accurately estimated. In a subsystem generated by a subarray, when the SNR is low, the noise seriously disturbs the detection of YOLO, thereby causing an extremely high false alarm rate, as illustrated in Fig.~\ref{Fig:SchemesL}. Most of the paths estimated by the alternative scheme are fake paths. With the increase in SNR, the probability of false alarm decreases. Nevertheless, the overhead amount still quadruples that of the proposed scheme when SNR $=10$ dB.

\begin{figure}
  \centering
  \includegraphics[scale=0.55]{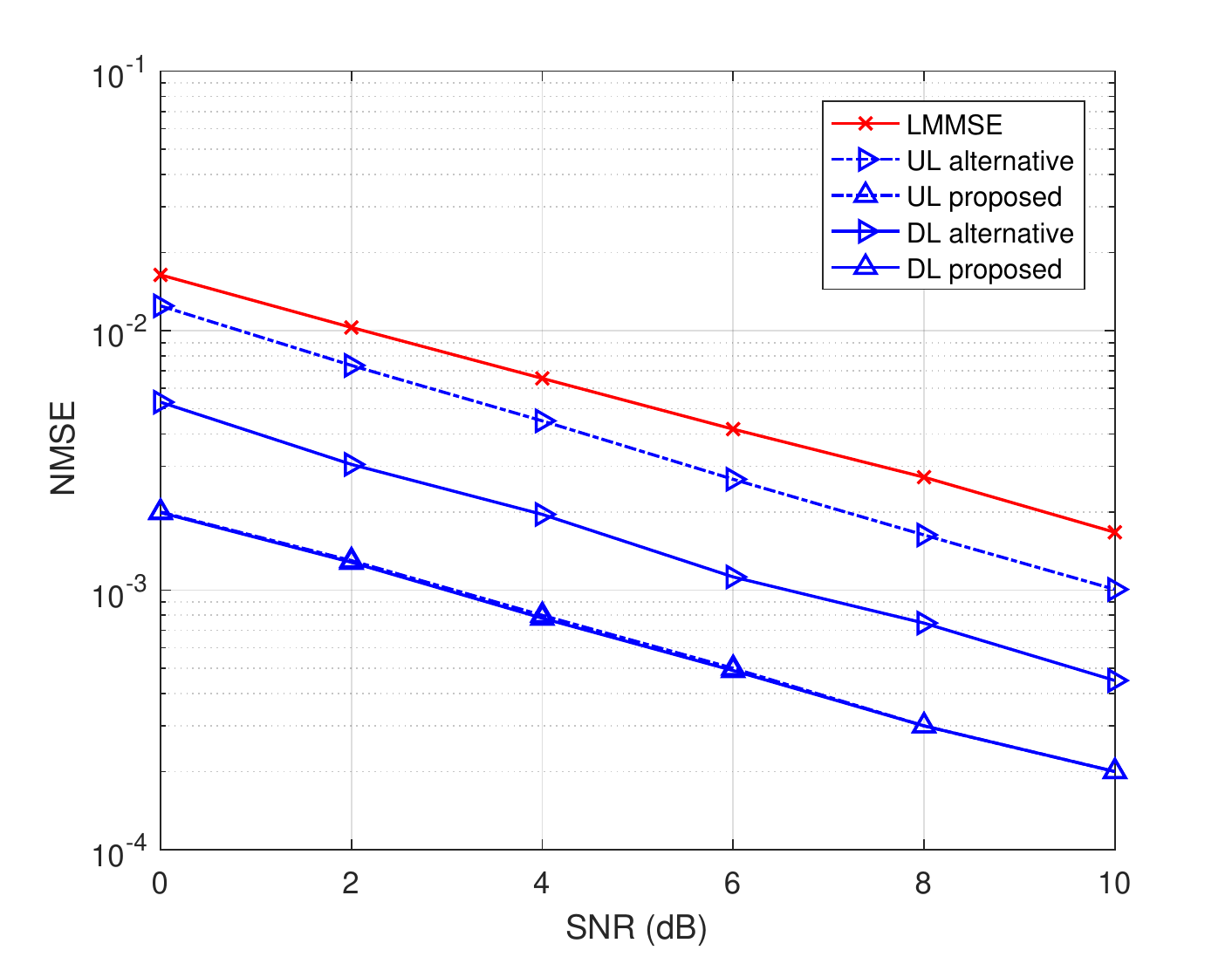}
  \caption{NMSEs of the proposed and the alternative schemes.} \label{Fig:SchemesNMSE}
\end{figure}

For the alternative scheme, with less measurements in each subsystem, the estimation accuracy of the angles and delays of actual paths is lower than that of the proposed scheme. Therefore, the NMSE of the uplink channel reconstructed by the alternative scheme is worse than that of the proposed scheme, as shown by the curve labeled as ``UL alternative'' in Fig.~\ref{Fig:SchemesNMSE}. Moreover, the large amount of fake paths causes the alternative scheme to outperform the LMMSE method by integrating these paths together, thereby compensating the estimation error of actual paths and achieving a high global accuracy. When reconstructing the downlink channel, the amount of downlink training and feedback overhead cost by the downlink gain estimation module of the alternative scheme is much larger than that of the proposed scheme. The alternative scheme has a good NMSE performance in reconstructing the downlink channel. Nevertheless, the NMSE of the alternative scheme is still inferior to that of the proposed scheme. The proposed scheme harvests the multi-subarray gain and achieves almost equivalent NMSE performance in reconstructing the uplink and downlink channels, demonstrating the accuracy of frequency-independent parameter estimation. Moreover, the NMSE performance of the two schemes under non-stationary condition is between that under the stationary conditions of $M=N=32$ and $M=N=128$. This phenomenon is in accordance with the assumption of the lower and upper bounds in Fig.~\ref{Fig:YOLOvsNOMP}.

\begin{figure}
  \centering
  \includegraphics[scale=0.55]{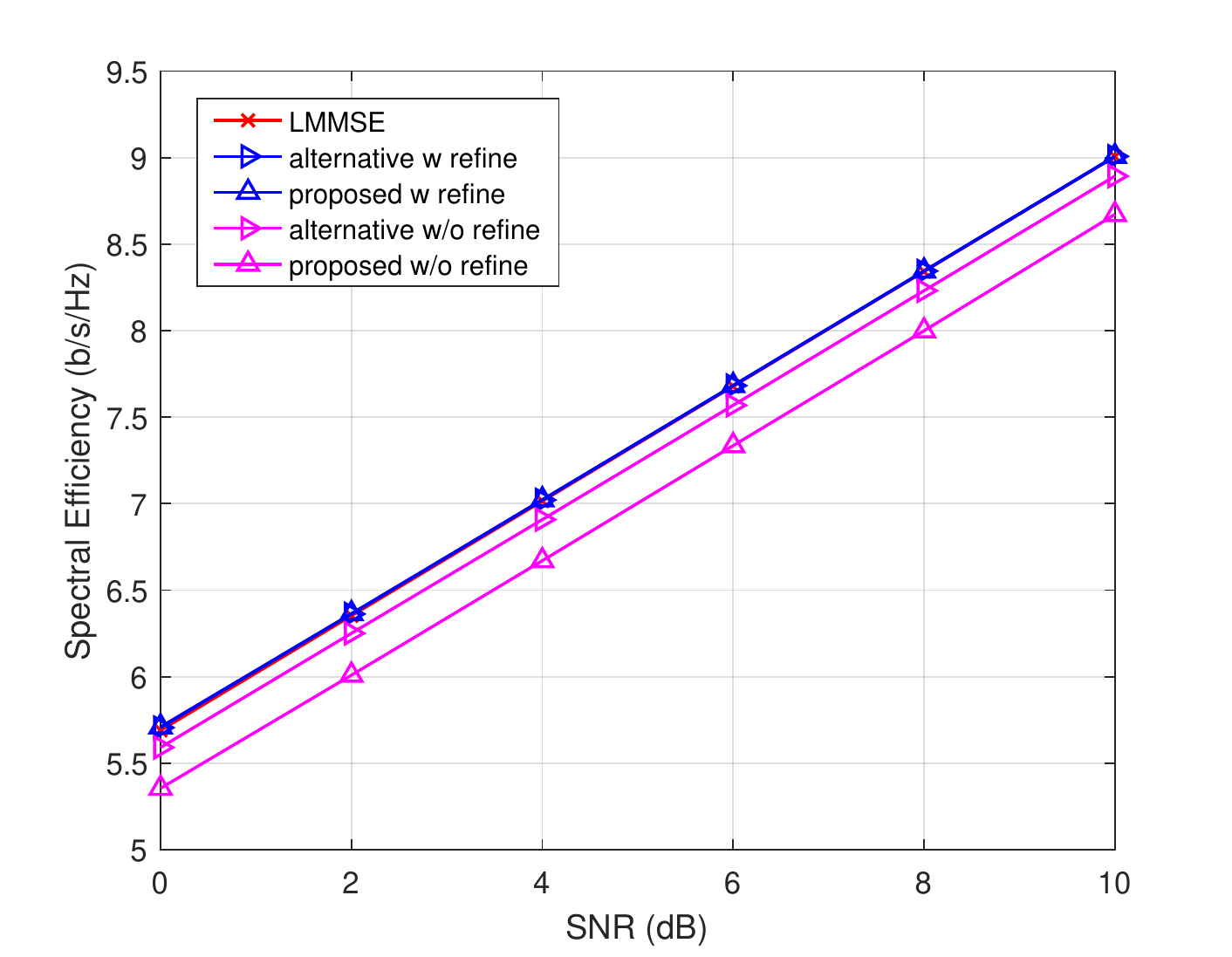}
  \caption{Spectral efficiency of the proposed and the alternative schemes.} \label{Fig:SchemesSE}
\end{figure}

We further compare the performance of the two schemes in practical systems. The spectral efficiency in the downlink is evaluated under the condition of maximal ratio transmitting. The signal received by the user can be expressed as
\begin{equation}\label{Eq:DLsignal}
{\bf r} = \sqrt{P} {\rm diag}\left\{{\bf H}^{\rm dl} \frac{{\hat{\bf H}}^{{\rm dl}H}}{\|{\hat{\bf H}}^{{\rm dl}H}\|}\right\} {\bf x} + {\bf z}^{\rm dl},
\end{equation}
where ${\bf r}\in\mathbb{C}^{N\times 1}$ with the $n$th entry as the received signal on the $n$th subcarrier, ${\bf x}\in\mathbb{C}^{N\times 1}$ is the transmitted signal across all subcarriers, satisfying $\mathbb{E}\{{\bf x}{\bf x}^H\} = {\bf I}$, and ${\bf z}^{\rm dl}\in\mathbb{C}^{N\times 1}$ is the noise whose elements are i.i.d. with zero mean and unit variance. When perfect downlink CSI is available at the user, the spectral efficiency can be calculated as
\begin{equation}\label{Eq:rate}
{\rm SE} = \mathbb{E}\left\{\frac{1}{N}\sum_{n=1}^N \log_2\left(1+\left|\left[{\bf H}^{\rm dl} \frac{{\hat{\bf H}}^{{\rm dl}H}}{\|{\hat{\bf H}}^{{\rm dl}H}\|}\right]_{n,n}\right|^2\right) \right\}.
\end{equation}
Fig.~\ref{Fig:SchemesSE} illustrates the Monte-Carlo results of the spectral efficiency. When the refiner is applied, the proposed scheme, the alternative scheme, and the LMMSE method have nearly the same spectral efficiency because their NMSEs are lower than $10^{-2}$. LMMSE requires 128 OFDM symbols for downlink training and feeds back $128\times 128$ complex numbers, whereas the proposed scheme only costs 1 to 10 OFDM symbols for downlink training and feeds back 1 to 10 complex numbers. If the refiner is absent, then the proposed scheme achieves relatively lower spectral efficiency than the alternative efficiency because of the smaller number of estimated paths, as well as the greatly reduced amount of downlink training and feedback overhead. On the other hand, although the NMSE performance is poor, the spectral efficiency is not badly impacted without the refinement module. Therefore, directly applying the learning-based estimates of parameters is acceptable in single user systems.

\section{Conclusion}\label{Sec:conclusion}

This study considered the FDD non-stationary massive MIMO system and proposed a deep learning-based scheme to reconstruct the downlink channel. Two key problems on the processing time and the non-stationary identification were successfully tackled by YOLO. The proposed downlink channel reconstruction scheme was designed to function in five modules given the power of YOLO. The visibility regions were detected by the non-stationary identifier, and estimation accuracy of angles and delays was improved by the angle and delay refiner. Moreover, the reduced case for stationary systems was discussed, and an alternative scheme for non-stationary systems was further analyzed. The numerical results verified the efficiency of the proposed scheme, and demonstrated that the NMSE was superior to that of the NOMP-based scheme in the FDD stationary massive MIMO systems.

\section{Appendix}\label{Sec:appendix}

\subsection{Proof of Property \ref{Theo:lightSpot1}}
The image of uplink pilots is generated by ${\tilde{\bf Y}}^{\rm ul}$, which is further obtained from ${\bar{\bf Y}}^{\rm ul}$. The $(m,n)$th entry of ${\bar{\bf Y}}^{\rm ul}_k$ is expressed as
\begin{equation}\label{Eq:appendix11}
[{\bar{\bf Y}}^{\rm ul}]_{m,n} = \sum_{l=1}^{L} \alpha_l \kappa_{{\rm a},l} \kappa_{{\rm t},l},
\end{equation}
where
\begin{equation}\label{Eq:appendix12}
\kappa_{{\rm a},l} = {\bf a}^H \left(-\frac{m}{\gamma_{\rm a}M}\right) \left({\bf a}(\Theta_l) \odot {\bf p}(\Phi_l) \right),
\end{equation}
and
\begin{equation}\label{Eq:appendix13}
\kappa_{{\rm t},l} = {{\bf q}^T(\Gamma_l){\bf q} \left(-\frac{n}{\gamma_{\rm t}N}\right)}.
\end{equation}
We initially derive the expression of $\kappa_{{\rm a},l}$. In accordance with \eqref{Eq:avec} and \eqref{Eq:pvec}, \eqref{Eq:appendix12} can be expressed by
\begin{equation}\label{Eq:appendix14}
\kappa_{{\rm a},l} = e^{j2\pi m_{l,{\rm start}} \left(\Theta_l+\frac{m}{\gamma_{\rm a}M}\right)} +\cdots+ e^{j2\pi m_{l,{\rm end}}\left(\Theta_l+\frac{m}{\gamma_{\rm a}M}\right)},
\end{equation}
where $m_{l,{\rm start}} = (s_{l,{\rm start}}-1)M/S+1$ and $m_{l,{\rm end}} = s_{l,{\rm end}}M/S$. Utilizing the feature of geometric progression, we can further express \eqref{Eq:appendix14} by
\begin{equation}\label{Eq:appendix15}
\begin{aligned}
&\kappa_{{\rm a},l} = \\
&\frac{1-e^{j2\pi\left(m_{l,{\rm end}}-m_{l,{\rm start}}+1\right)\left(\Theta_l+\frac{m}{\gamma_{\rm a}M}\right)}} {1-e^{j2\pi\left(\Theta_l+\frac{m}{\gamma_{\rm a}M}\right)}}  e^{j2\pi m_{l,{\rm start}}\left(\Theta_l+\frac{m}{\gamma_{\rm a}M}\right)}.
\end{aligned}
\end{equation}
In addition,
\begin{equation}\label{Eq:appendix16}
{1-e^{j2\pi\left(\Theta_l+\frac{m}{\gamma_{\rm a}M}\right)}} = e^{j\pi\left(\Theta_l+\frac{m}{\gamma_{\rm a}M}\right)} \sin\left(\pi\left(\Theta_l+\frac{m}{\gamma_{\rm a}M}\right) \right).
\end{equation}
Thereafter, we can obtain the module of $\kappa_{{\rm a},l}$ as
\begin{equation}\label{Eq:appendix17}
|\kappa_{{\rm a},l}| = \frac{\sin\left(\pi\left(m_{l,{\rm end}}-m_{l,{\rm start}}+1\right)\left(\Theta_l+\frac{m}{\gamma_{\rm a}M}\right) \right)} {\sin\left(\pi\left(\Theta_l+\frac{1}{\gamma_{\rm a}M}\right) \right)},
\end{equation}
which is the sinc function shown in Fig.~\ref{Fig:CoordinateSystems}(b). In accordance with \eqref{Eq:appendix17}, $|\kappa_{{\rm a},l}|$ achieves its maximal value, i.e., $m_{l,{\rm end}}-m_{l,{\rm start}}+1$, when $\Theta_l+{m}/({\gamma_{\rm a}M})=0$. The center of the $l$th dark spot has the maximal value. Thus, the vertical coordinate of the dark spot center is
\begin{equation}\label{Eq:appendix18}
y=-\frac{m}{\gamma_{\rm a}M}=\Theta_l.
\end{equation}

Similarly, we calculate the module of $\kappa_{{\rm t},l}$ as
\begin{equation}\label{Eq:appendix19}
|\kappa_{{\rm t},l}| = \frac{\sin\left(\pi N \left(\Gamma_l-\frac{n}{\gamma_{\rm t}N} \right)\right)} {\sin\left(\pi \left(\Gamma_l-\frac{n}{\gamma_{\rm t}N} \right)\right)}.
\end{equation}
In accordance with \eqref{Eq:appendix19}, $\kappa_{{\rm t},l}$ achieves its maximal value, i.e., $N$, when $\Gamma_l-{n}/({\gamma_{\rm t}N})=0$. Thus, the horizontal coordinate of the dark spot center of the $l$th cross-style pattern is
\begin{equation}\label{Eq:appendix20}
x=\frac{n}{\gamma_{\rm t}N}=\Gamma_l.
\end{equation}

\subsection{Proof of Property \ref{Theo:lightSpot2}}

Given that $m_{l,{\rm end}}-m_{l,{\rm start}}+1 = (s_{l,{\rm end}}-s_{l,{\rm start}}+1)M/S$, \eqref{Eq:appendix17} can be further expressed by
\begin{equation}\label{Eq:appendix21}
|\kappa_{{\rm a},l}| = \frac{\sin\left(\pi\frac{M}{S}\left(s_{l,{\rm end}}-s_{l,{\rm start}}+1\right)\left(\Theta_l+\frac{m}{\gamma_{\rm a}M}\right) \right)} {\sin\left(\pi\left(\Theta_l+\frac{1}{\gamma_{\rm a}M}\right) \right)}.
\end{equation}
\eqref{Eq:appendix21} shows that $\kappa_{{\rm a},l}$ achieves its minimum value, i.e., 0, when
\begin{equation}\label{Eq:appendix22}
\frac{M}{S}\left(s_{l,{\rm end}}-s_{l,{\rm start}}+1\right)\left(\Theta_l+\frac{m}{\gamma_{\rm a}M}\right) = q,
\end{equation}
where $q$ is an nonzero integer. The vertical coordinates of the white points along the vertical central line of the dark point are
\begin{equation}\label{Eq:appendix23}
y = -\frac{m}{\gamma_{\rm a}M} = \Theta_l+\frac{qh_l}{2},
\end{equation}
and $h_l$ is defined in \eqref{Eq:WidthHeight}. The vertical coordinate of the dark spot is $\Theta_l$; thus, the half-height of the dark spot is
\begin{equation}\label{Eq:appendix24}
\left|\left(\Theta_l-\frac{h_l}{2}\right)-\Theta_l\right|=\frac{h_l}{2}.
\end{equation}

Similarly, \eqref{Eq:appendix19} shows that $\kappa_{{\rm t},l}$ achieves its minimum value, i.e., 0, when
\begin{equation}\label{Eq:appendix25}
\Gamma_l-\frac{n}{{\gamma_{\rm t}N}} = \frac{qw_l}{2}.
\end{equation}
Accordingly, the half-width of the dark spot is $w_l/2$.

\subsection{Proof of Property \ref{Theo:PsApprox}}
If ${\tilde\Theta}_l\approx\Theta_l$ and ${\tilde\Gamma}_l\approx\Gamma_l$, then $P_{l,s}$ is approximated by $P_{l,s}\approx P^{\rm ul}\sum_{k=1}^L {\eta_{1,k}} +\eta_2$, where
\begin{equation}\label{Eq:appendix31}
\begin{aligned}
&\eta_{1,k} = \\ &\left|\alpha_k \left({\bf a}(\Theta_l)\odot{\bf p}(\{s\})\right)^H \left({\bf a}(\Theta_k)\odot{\bf p}(\Phi_k)\right){\bf q}^T(\Gamma_k) {\bf q}^*({\Gamma}_l)\right|^2,
\end{aligned}
\end{equation}
and
\begin{equation}\label{Eq:appendix32}
\eta_{2} = \left|\left({\bf a}(\Theta_l)\odot{\bf p}(\{s\})\right)^H {\bf Z}^{\rm ul} {\bf q}^*({\Gamma}_l)\right|^2.
\end{equation}
If $M/K$ is large, then $P_{l,s}\approx\mathbb{E}\{P_{l,s}\}$,
\begin{equation}\label{Eq:appendix33}
\mathbb{E}\{\eta_{1,k}\} =
\begin{cases}
P^{\rm ul}|\alpha_l|^2{M^2N^2}/{S^2}, & \text{if $k=l$ and $s\in\Phi_l$}, \\
0, & \text{if $k\ne l$},
\end{cases}
\end{equation}
and $\mathbb{E}\{\eta_{2}\} \approx {MN}/{S}$ hold. Finally, we obtain \eqref{Eq:PsApprox} by applying \eqref{Eq:appendix32} and \eqref{Eq:appendix33}.

\end{document}